\documentclass[journal, compsoc]{IEEEtran}
\usepackage{amsmath,amsfonts}
\usepackage{algorithmic}
\usepackage{algorithm}
\usepackage{array}
\usepackage{textcomp}
\usepackage{stfloats}
\usepackage{url}
\usepackage{verbatim}
\usepackage{graphicx}
\usepackage{cite}
\usepackage[colorlinks=true,linkcolor=blue,citecolor=blue,urlcolor=blue]{hyperref}
\usepackage{cleveref}
\usepackage{ulem}
\usepackage{orcidlink}
\usepackage{makecell}
\usepackage{comment}
\usepackage{xspace}
\usepackage{svg}
\usepackage{gensymb}
\usepackage{subcaption}
\begin{document}
\def\UrlBreaks{\do\/\do-\do.}

\newcommand{\myemph}[1]{\textit{#1}}
\newcommand{\etal}{\myemph{et al.}\xspace}
\newcommand{\SoS}{\myemph{Science On a Sphere}\xspace}


\title{Augmenting a Large Language Model\\
with a Combination of Text and Visual Data\\ for Conversational Visualization\\ of Global Geospatial Data}

\author{Omar Mena~\orcidlink{0009-0002-6564-0175}, Alexandre Kouyoumdjian~\orcidlink{0009-0005-0915-369X}, Lonni Besançon~\orcidlink{0000-0002-7207-1276}, Michael Gleicher~\orcidlink{0000-0003-3295-4071}, Ivan Viola~\orcidlink{0000-0003-4248-6574} and Anders Ynnerman~\orcidlink{0000-0002-9466-9826}

\thanks{Omar Mena, Alexandre Kouyoumdjian and Ivan Viola are affiliated with the King Abdullah University of Science and Technology
(KAUST), Saudi Arabia. 
E-mail: \{omar.mena\,$|$\,alexandre.kouyoumdjian\,$|$\,ivan.viola\}\\*@kaust.edu.sa}
\thanks{Michael Gleicher is with the Department of Computer Sciences, University of Wisconsin-Madison, United States of America. 
   E-mail: gleicher@cs.wisc.edu}
\thanks{
Lonni Besançon and Anders Ynnerman are with the Department of Science and Technology, Linköping University, Sweden.
  E-mail: \{lonni.besancon\,$|$\,anders.ynnerman\}@liu.se
}
\thanks{Manuscript received April 19, 2021; revised August 16, 2021.}}



\maketitle

\begin{abstract}

We present a method for augmenting a Large Language Model (LLM) with a combination of text and visual data to enable accurate question answering in visualization of scientific data, making conversational visualization possible. LLMs struggle with tasks like visual data interaction, as they lack contextual visual information. We address this problem by merging a text description of a visualization and dataset with snapshots of the visualization. We extract their essential features into a structured text file, highly compact, yet descriptive enough to appropriately augment the LLM with contextual information, without any fine-tuning. This approach can be applied to any visualization that is already finally rendered, as long as it is associated with some textual description.

We apply our method on geospatial data from the Science On a Sphere (SoS) project. We demonstrate that by simply conversing with the augmented LLM in any common language, users can choose the dataset, the viewing direction, or retrieve contextual information from the dataset’s description or the LLM’s vast general knowledge. We evaluate our approach in a user study, showing it to be accurate, and well-liked by participants.
\end{abstract}


\section{Introduction}

Though visualization may be of great help in understanding scientific findings, without proper explanation and contextualization, people (especially nonexperts) can struggle to understand what they are seeing~\cite{smith2011aesthetics}. Smith \etal specifically mention that ``explanations should include information about colors used, size, scale, and location of the object''. For example, a viewer looking at a map of sea temperatures might wonder about odd color patterns near the equator. A human facilitator can focus the view on this area and explain that these patterns represent sea temperature variations, which are caused by upwelling of cooler subsurface waters. Such a response requires a coherent combination of knowledge about the visualization, the underlying data, and the general scientific background behind it. It also requires the ability to form a focused and informative response based on this combination of knowledge. Thus, until now, such conversational interactions have only been possible with skilled human experts.

However, human experts are a scarce and expensive resource, which makes automating such interactions attractive. Indeed, in \myemph{conversational visualization}, the viewer engages in a dialogue about the visual material, fostering better understanding and greater engagement. In the example above, the system would automatically focus the view and provide the response. With automation, this type of interaction can be brought to broader audiences, provided the system can answer questions about a visualization by making use of the visualization itself, its background, and the underlying science. While large language models excel at interpreting questions and referring to general scientific knowledge, they are unable to refer to a visualization while discussing it, because they cannot ``see'' it. Indeed, LLMs can only process text, they are unable to read visual information of any kind on their own, without assistance from a visual model, or through other means, such as advance visual tokenizers. This limitation in prior efforts is at least partially due to the lack of any established methods for augmenting an LLM with visual information, let alone combining this visual information with text augmentation data in a coherent manner. This problem of ``LLM blindness'' is a core challenge in creating a conversational visualization agent.

\begin{figure}[htb!]
    \centering
    \includegraphics[width=\columnwidth]{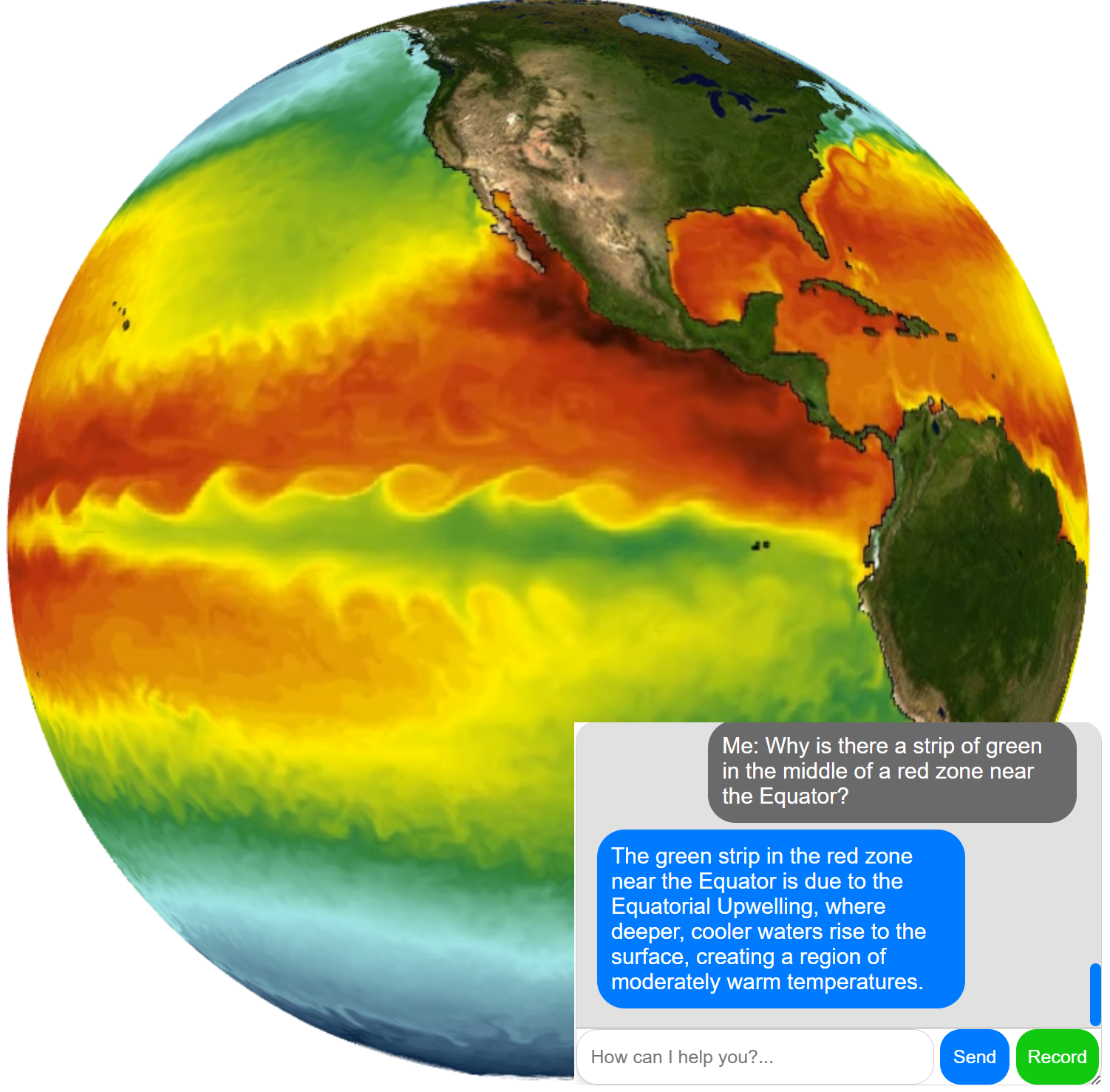}
    \caption{In response to a question about something the user observed on the visualization, the system combines visual information, context information about the specific dataset and the colors used, and its general knowledge of climatology and sea surface temperatures to provide an accurate and informative answer.}
    \label{fig:intro_example}
\end{figure}

In this paper, we present a new method that addresses the LLM blindness problem, enabling true conversational visualization. The key to our approach is to pre-process the visualization and its metadata to generate \myemph{structured} augmentation data that allow the question-answering LLM to refer to the visualization, but also to additional, specific information about the visualization, and to the LLM's general scientific knowledge. Our approach relies on prompt engineering without any fine-tuning or retraining, making it LLM agnostic, cheap and easy to deploy. We demonstrate this approach on global geospatial data from the \SoS~\cite{goldman2010science} (SoS) project.

\begin{figure}[ht]
    \centering
    \includegraphics[width=\columnwidth]{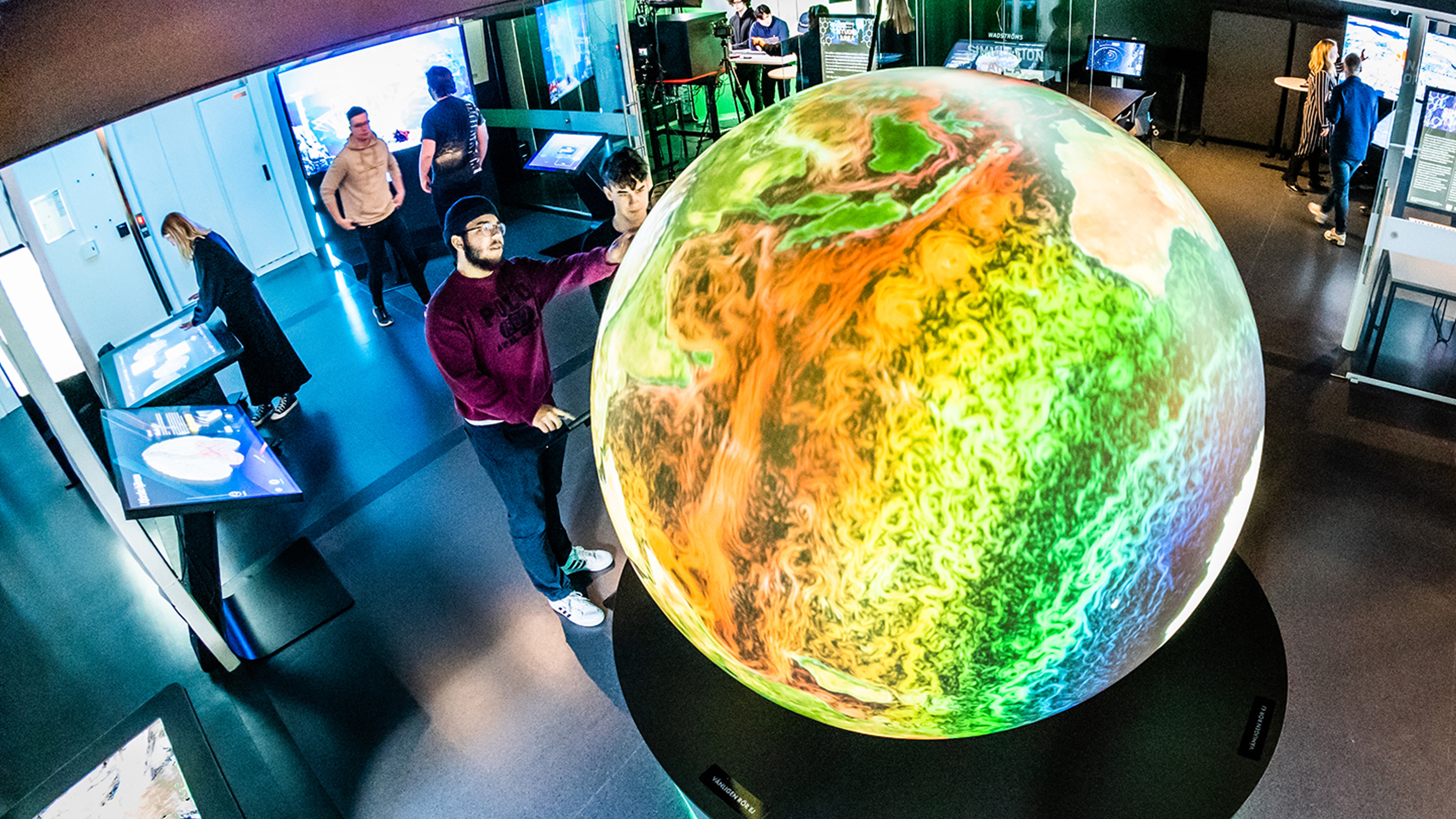}
    \caption{The \myemph{TellUs} sphere at the Norrköping Visualization Center C. It provides the framework for the development of technologies enabling a talking visioverbal planet for outreach at science centers and classroom settings.}
    \label{fig:tellus_sphere}
\end{figure}

This conversational visualization challenge is particularly relevant in the context of the \myemph{TellUs}\footnote{\footnotesize{\url{https://visualiseringscenter.se/en/research-program/tellus/}}} initiative~\cite{tellus} at the Norrköping Visualization Center C. The overall mission of this initiative is to increase the interest in science, technology, engineering, and math among visitors to science centers and schoolchildren in classrooms. The underpinning idea is to use novel visualization and interaction techniques together with advances in LLMs to produce a talking planet: \myemph{TellUs}, shown on \cref{fig:tellus_sphere}. This paper addresses the LLM blindness problem, providing one core technical contribution towards the \myemph{TellUs} vision. In section~\ref{sec:design} we elaborate on the requirements of the method presented here, given the need of the \myemph{TellUs} initiative (or of similar endeavors) to make LLMs capable of answering questions about visual data.

We present a proof of concept of this method: a conversational agent that can discuss datasets from the \SoS project, which are ``baked visualizations'' -- pre-rendered animated visual representations of the data that do not require real-time computation, produced by the National Oceanic and Atmospheric Administration\footnotemark{} (NOAA). This NOAA project aims to display such datasets on a spherical display to visualize scientific, planetary data and communicate different Earth phenomena visualizations to the public. These datasets are \myemph{only} pre-rendered videos, and do not contain any of the data used the generate the animations in the first place, although they do come with text descriptions.

Although the proof of concept demonstrated here supports spherical displays, it can also render a 3D sphere on a conventional display or an XR device~\cite{doerner2022virtual}, or even a smartphone. This ability to run on widespread devices with limited computational capabilities ensures broad accessibility, especially since it enables conversational and visual interaction with an LLM in a large number of languages. Experiments with users show that the system can successfully provide conversational support to viewers. These experiments also provide a corpus of example questions that allow us to quantify the correctness of an LLM's outputs once it is augmented with our method.

The main contribution of the paper lies in the augmentation of the LLM with dataset-specific, space-efficient metadata generated from both the textual description of the given dataset that is provided with it and extraction of textual information from rendered images of the dataset, using vision models such as OpenAI's VISION. We also contribute a proof of concept for this augmentation method, applied to geospatial data. We evaluate the correctness of the augmented LLM compared to two baselines featuring augmentation without the use of a structured approach. We also conduct an evaluation of the user experience with our proof of concept.

\footnotetext{\footnotesize{\url{https://www.noaa.gov}}}

\section{Related Work}
Significant research has been conducted on generating or controlling visualizations through natural language, mostly for 2D visualizations: plots, maps and graphs. Usually, the natural language processing involves specifically trained or fine-tuned models of various kinds. For an automated system to act as a facilitator, these capabilities are required. But comparatively little work has been done on models capable of ``seeing'' visualizations provided as input, and answering questions about them.

Indeed, LLM augmentation efforts have mostly focused on non-visual data, such as the \myemph{LLM-Augmenter}~\cite{peng2023check} system, designed to ground LLM-generated responses in verified information, such as task-specific databases, to improve accuracy and reduce hallucinations. By extending the \myemph{NL4DV} toolkit~\cite{narechania2020nl4dv}, the \myemph{Chat2Vis} system~\cite{maddigan2023chat2vis} generates data visualizations from conversational interaction with an LLM, through prompt-engineering and LLM-generated Python scripts, but cannot generate 3D scientific visualizations. \myemph{ADVISor}~\cite{liu2021advisor} is similar, but relies on a slightly different approach, using the \myemph{BERT}~\cite{devlin2018bert} language model to encode both natural language queries and tabular data into vectors, and generate complex data visualizations from them. The \myemph{VisEval} benchmark~\cite{chen2024viseval} evaluates LLMs' ability to generate visualizations from natural language. 
Mitra \etal\cite{mitra2022facilitating} also extended the \myemph{NL4DV} toolkit to support multi-turn dialogue, allowing users to progressively define and refine the visualization they want over successive natural language queries. The system converts them to Vega-lite~\cite{satyanarayan2016vega} specifications. It also identifies ambiguities in user requests and asks for clarification when needed. The VIST5~\cite{voigt2023vist5} system also generates Vega-lite specifications, but places greater emphasis on adaptability, allowing it to transfer functionalities across different applications. The solution developed by Hong \etal\cite{hong2023conversational} is similar, but includes support for ``multithreaded'' conversations, i.e., conversations spanning multiple branches represented explicitly. Multi-turn dialogue is essential to conversational interaction, whether it is to generate visualizations or to understand existing ones.

Jia \etal~\cite{jia2023voice} have already enabled conversational visualization and manipulation of scientific data with the \myemph{VOICE} system, designed for 3D models of biomolecular structures. It is an oracle that relies on a two-level architecture that involves a fine-tuned GPT3.5 model. This fine-tuning and few-shot prompt engineering allows the model to properly convert user queries into machine-readable commands that enable interaction with the system, but does not augment it with information (visual or otherwise) specific to the data being visualized. Thus, questions about the visualization are sent to GPT-4, and it relies on its existing training to provide explanations about it, without any ability to ``see'' the models.

But efforts to extract visual information in question-answering systems have also been made, even before the advent of LLMs, for example with DVQA~\cite{kafle2018dvqa}, which focuses on information extraction and question answering about bar charts, where the \emph{SANDY} model uses a local dictionary to dynamically encode chart-specific words. Likewise, Masry \etal~\cite{masry2022chartqa} devised a system to use visual information from charts. They augmented \myemph{TaPas}, a BERT-based transformer~\cite{vaswani2017attention} model designed for structured data with a vision transformer~\cite{dosovitskiy2020image} to create \myemph{VisionTaPas}, thanks to a cross-modality encoder~\cite{li2021align}. This allows their solution to answer questions about charts based on their visual elements. In the same domain, Bendeck \etal~\cite{bendeck2024empirical} demonstrated that GPT-4 VISION model, when provided with well-structured inputs, improves its ability to interpret visualizations and provide accurate, context-aware responses. Similarly, Feng \etal~\cite{feng2023geoqamap} investigated a similar approach for geographic data, but took the problem from the other end: their \myemph{GeoQAMap} is meant to answer geographic questions formulated in natural language. It translates them into \myemph{SPARQL}~\cite{harris2013sparql} queries, retrieves geospatial information from Wikidata,\footnotemark and generates visual answers in the shape of annotated and interactive maps. However, it cannot extract visual information from existing maps. Chang \etal~\cite{chang2022mapqa} addressed this using a ``visual multi-output data extraction model''. It extracts structured data from choropleth maps, but only from choropleth maps, and it can only answer questions, not engage in proper conversational interaction with context retention or multi-turn conversations, as it lacks a sophisticated language model. 
Bursztyn \etal~\cite{chartsAsText} demonstrated that using chart specifications in text form allows LLMs to outperform vision-based models like OpenAI's VISION in chart-grounded question answering (CQA) tasks. Their text-based approach excels at generating accurate explanations and answers, highlighting the benefits of text over vision models in these contexts.

While we are not aware of any efforts to apply this to visualization, Dialogue State Tracking~\cite{young2013pomdp} aims to maintain a \myemph{belief state} about the user's objectives and actions, and the overall state of the conversation. It summarizes information in slot-value pairs, along with their probabilities. This approach helps managing complex conversations with many inputs over time.

\footnotetext{\footnotesize{\url{https://www.wikidata.org}}}

\section{Design Requirements}
\label{sec:design}

The general goal of the \myemph{TellUs} initiative is to produce a talking planet named \myemph{TellUs} that is capable of providing visioverbal interactions with visitors to science centers. To achieve this goal, it is particularly important that the prototype that we aim to develop be able to answer questions such as: ``What does the green color mean?'', ``What is the red area next to Japan?'', ``What do these symbols represent?'', ``What are the red zones?'', or ``How do these blue points relate to the green ones?''. 

The \myemph{TellUs} initiative aims to use novel visualization and interaction techniques together with advances in LLMs to achieve its general goal. So, such queries to be accurately answered, we need to satisfy the following three key design requirements:

\begin{itemize}

\item \textbf{DR1--Visual awareness}: The method should enable the LLM to explain each dataset to users: to tell them what they see in the visualization, what the different colors or symbols mean, how they relate to one another, and how they should interpret the visuals. This information should be highly accurate, both visually and scientifically. In other words, the LLM should be able to identify and locate all important visual elements, but also correctly relate them to their scientific meaning. Only then are users guaranteed to obtain the right answer to their scientific questions. 

\item \textbf{DR2--General contextual information}: The method should not prevent the LLM from accessing its existing contextual information about any dataset, far beyond the metadata included with the dataset. For instance, when visualizing a dataset presenting historical data about tsunamis, the LLM should be able to explain what a tsunami is, how it forms, how dangerous it is, etc., even if the dataset's description lacks this information. Therefore, we should not rely on a specifically trained model but instead ensure that we can use a general language model such as GPT-4o. 

\item \textbf{DR3--Extensibility}: It is important to provide a solution that is extensible. The method should work with new datasets with minimal effort, without any need to fine-tune or retrain any model, because of the complexity of such tasks. So, we aim to provide a solution that can easily be extended to new datasets with simple and inexpensive efforts from humans, even without specific expertise in LLMs.

\end{itemize} 

Here, we focus on these core design requirements for augmenting an LLM with visual and contextual information about scientific visualizations. But in designing our proof of concept, we did keep additional requirements in mind, such as ease of access through a web-based application, as well as keeping latencies low, and minimizing the financial cost of API calls to LLMs.


\section{Method for Visioverbal Augmentation}

\begin{figure}[htb!]
    \centering
    \includegraphics[width=\columnwidth]{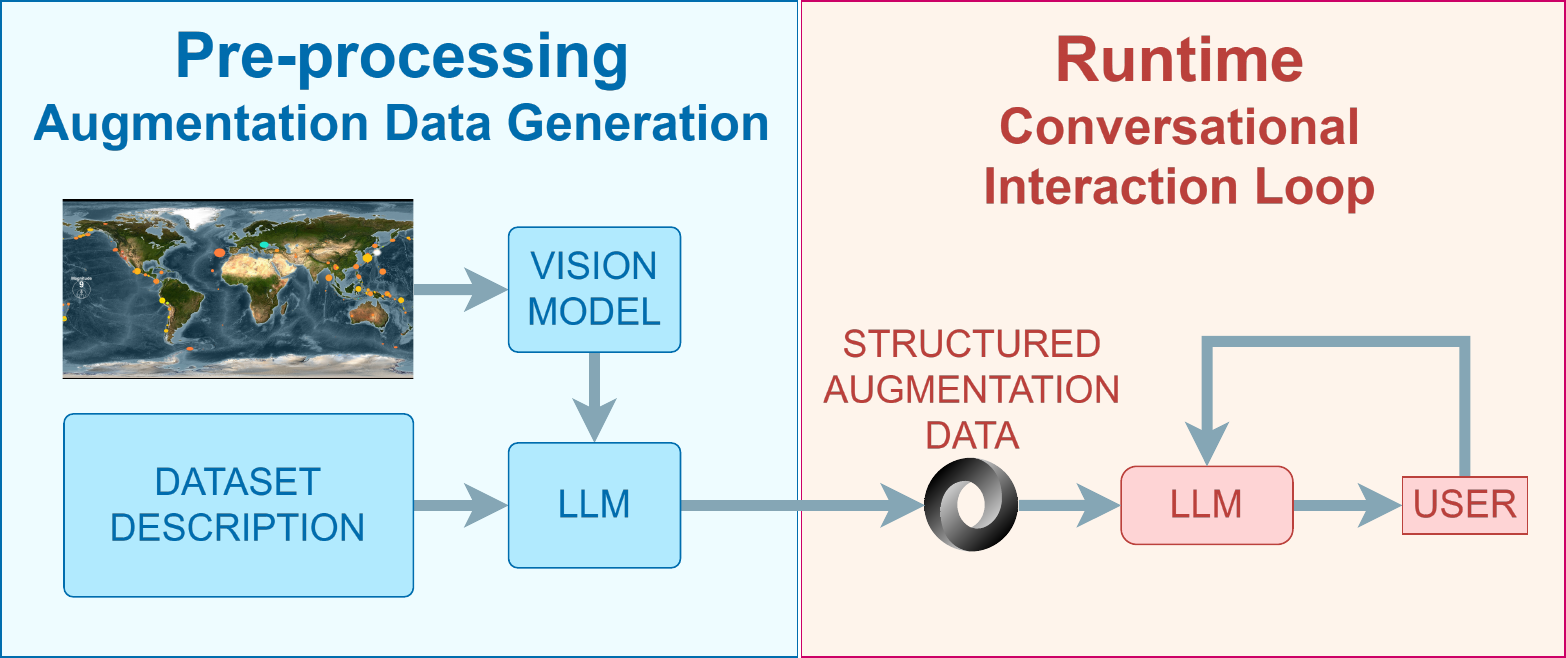}

    \caption{Simplified overview of our method for conversational interaction. Left: the LLM receives extracted information from the vision model and from the dataset's description, and generates structured augmentation data from these inputs. Right: at runtime, these augmentation data augment the LLM, providing it with the required information to respond to user queries.}
    \label{fig:simple_overview}
\end{figure}

The core idea in the method we introduce is to pre-process the available information about the visualization, including rendered frames from the visualization itself, plus its text description, into a \myemph{structured}, compact form that can be used as augmentation data by the LLM. A simplified overview of this process is shown on \cref{fig:simple_overview}. This method comprises two major parts: a pre-processing step, run once per dataset, and a runtime component, the main conversational interaction loop, which runs continuously while the system is in use. In the pre-processing stage, the vision model extracts information from the rendered frames, feeds the extracted information to an LLM, which also receives the dataset's description. From these inputs, the LLM generates structured augmentation data. At runtime, in the conversational interaction loop, these augmentation data are added to every prompt, which augments the LLM with all the information it requires, in a format it can parse and use reliably.

\begin{figure*}[ht]     
    \centering
    \includegraphics[width=\textwidth]{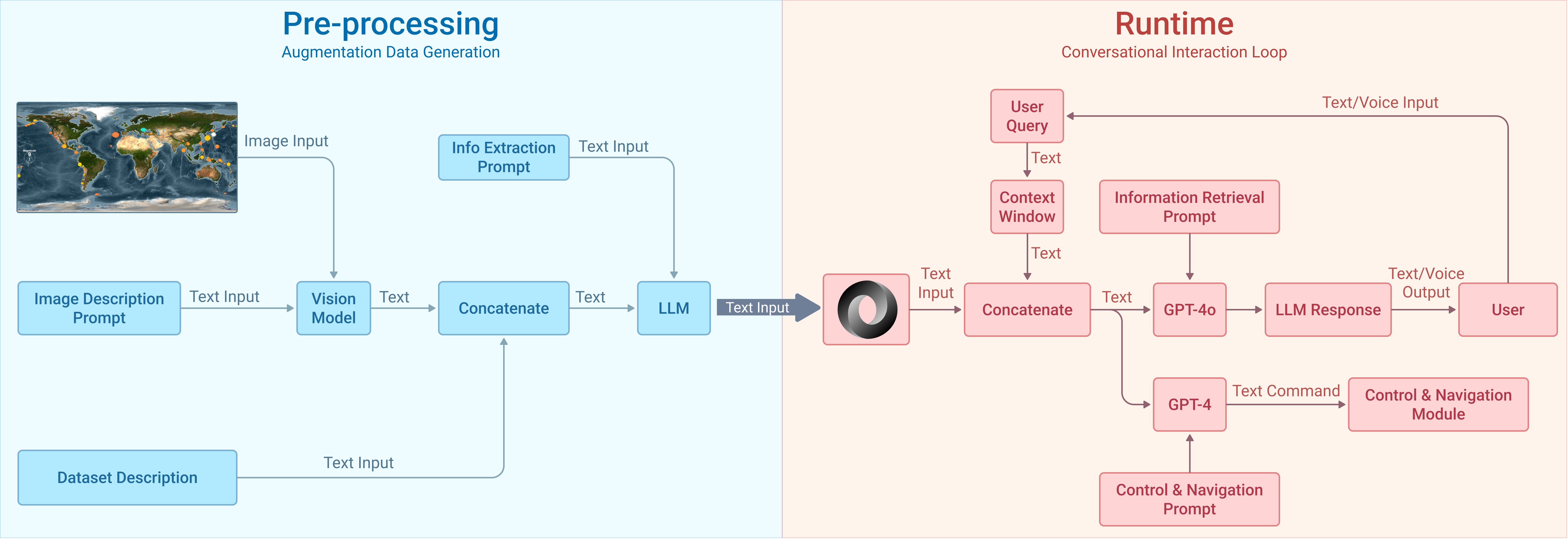}
    \caption{Detailed overview of our method for conversational visualization. \textbf{Left:} the pre-processing step executed just once per dataset upon its addition to our database. Images are fed to OpenAI's VISION model along with a prompt engineered to get the model to extract visual information from the images, and output it as text. This is concatenated with the text description of the dataset, and fed to GPT-4o with a prompt that instructs the model to extract information from this concatenated text into a structured JSON format. \textbf{Right:} the main interaction loop of our system. The user's query is added to the context window which contains all previous queries (up to 20), and concatenated with the JSON structure. This is then fed to both LLMs with appropriate prompts. The user's query is always fed to the GPT-4 model along with a prompt instructing it to look for a control or navigation command expressed in natural language, and process it if it is there. If it finds such a command, it converts it to a formal one that our system can parse and deterministically execute. If the query was in fact a request for information, then GPT-4o's response is presented to the user, who can then generate another query.}
    \label{fig:omar_earth_overview}
\end{figure*}

A complete overview of our method is shown on  \cref{fig:omar_earth_overview}. It provides further detail on both stages: pre-processing and runtime, and here we give a thorough explanation of the method it illustrates.

\subsection{Pre-processing: Augmentation Data Generation}
To satisfy both \textbf{DR1--Visual awareness} and \textbf{DR2--General contextual information}, our method must be able to provide answers augmented with a \myemph{combination} of text information about a given dataset \myemph{and} visual information about the visualization, plus the LLM's general scientific knowledge.

The pre-processing component is run when a scientific dataset is added to our collection of datasets. Its purpose is to extract information from each dataset's text description as well as from a number of frames sampled from the rendered visualization (an animated video of a world map), and to save this information as structured text files (in the JSON format). The use of a compact, structured format follows the same principles as Dialogue State Tracking~\cite{young2013pomdp}, but applied to a single large text input, rather than an actual dialogue. It leverages the fact that LLMs are trained on extensive datasets and possess strong pattern recognition and logical relational capabilities, and tend to perform better with structured inputs.

\subsubsection{Frame Sampling}
\label{sec:sampling}
The first step to generating such structured augmentation data is to sample rendered frames from the visualization. The number of frames processed is arbitrary---the more frames we choose to process, the greater the accuracy of feature extraction, but this also increases the cost and processing time for the vision model. We choose a default value of two, keeping processing time to around a minute, helping us satisfy \textbf{DR3–Extensibility}, with good accuracy. This value works well because while the visualizations in \SoS datasets do change over time, they often involve changes in size or position for visual elements, but very rarely their sudden appearance or disappearance. Thus, a couple of samples will usually capture all of the visual elements ever displayed on the map in the given dataset. For datasets where the visualization changes drastically across different frames, especially with elements disappearing or new ones appearing, greater sample counts may be necessary. Figure ~\ref{fig:tsunami_frames} shows an example of such a dataset.

\newcommand{\framewidth}{0.118\textwidth}
\begin{figure}[htb]
    \centering
    
    \includegraphics[width=\framewidth]{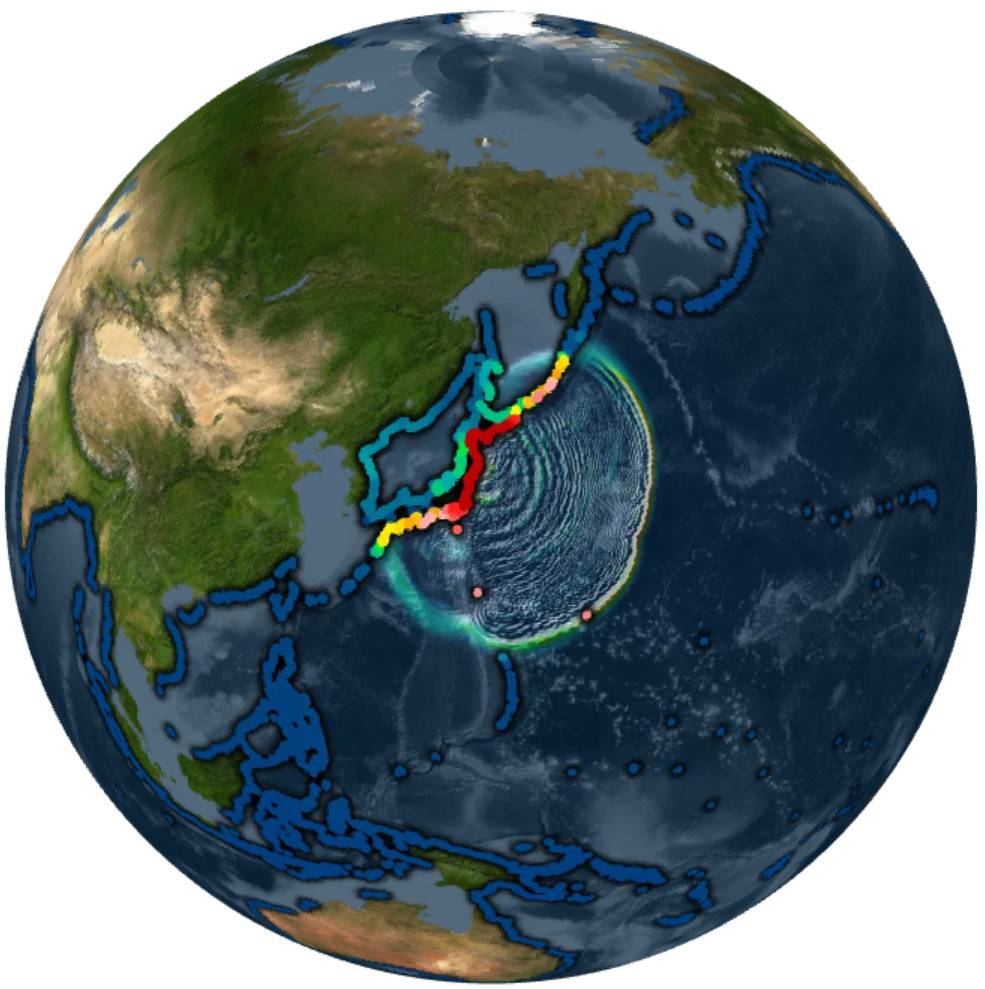}
    \includegraphics[width=\framewidth]{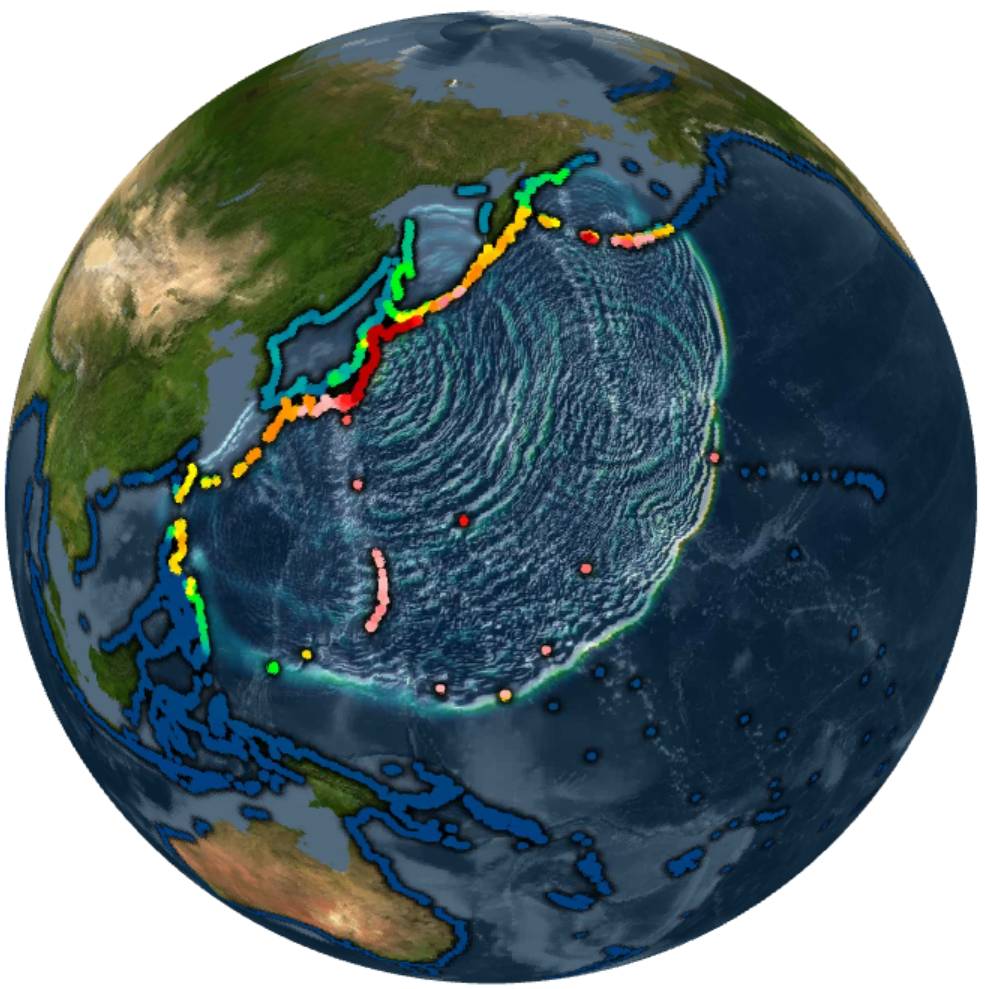}
    \includegraphics[width=\framewidth]{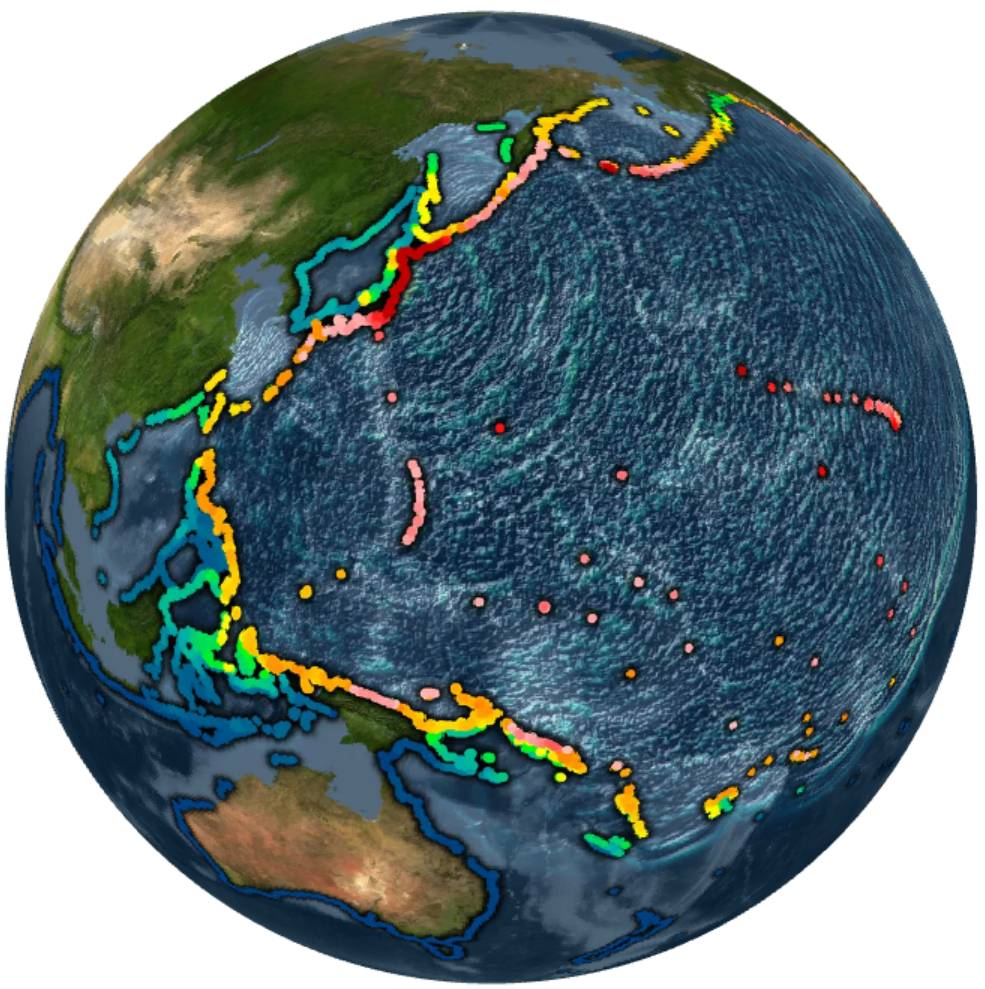}
    \includegraphics[width=\framewidth]{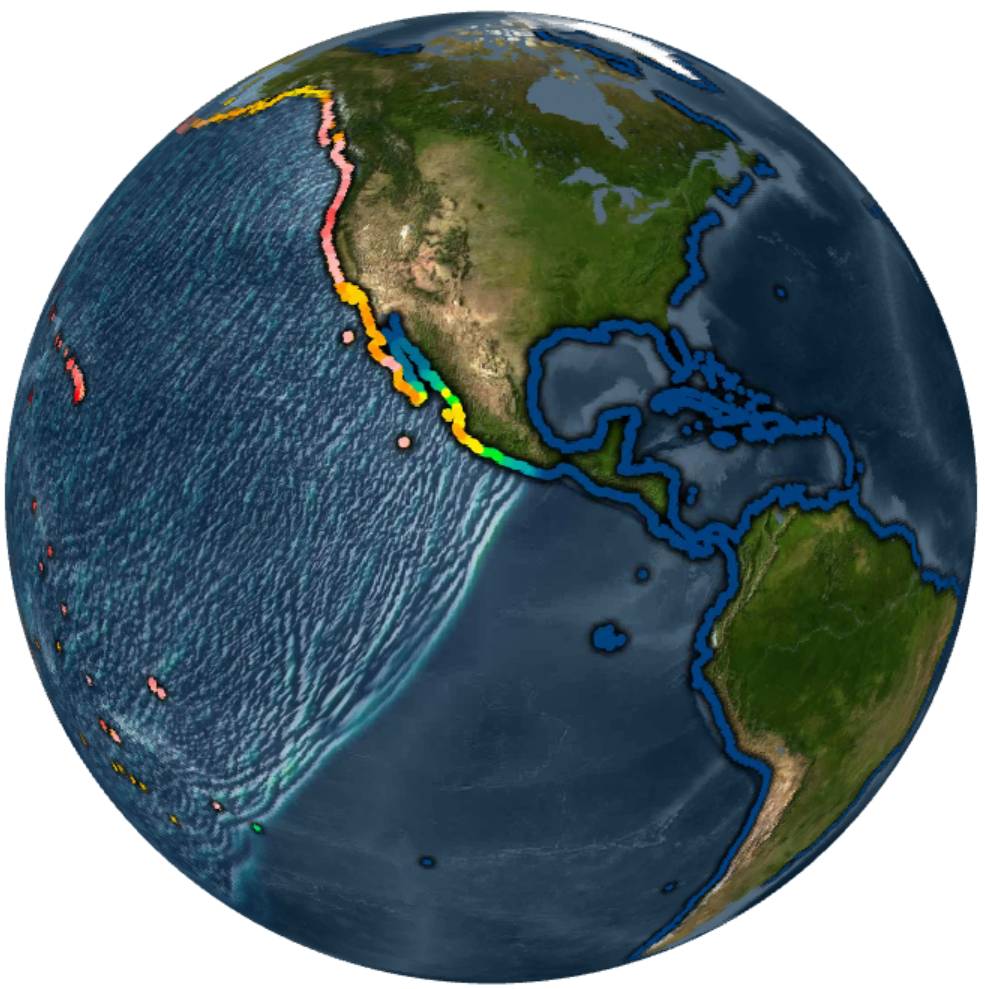}
    \includegraphics[width=\framewidth]{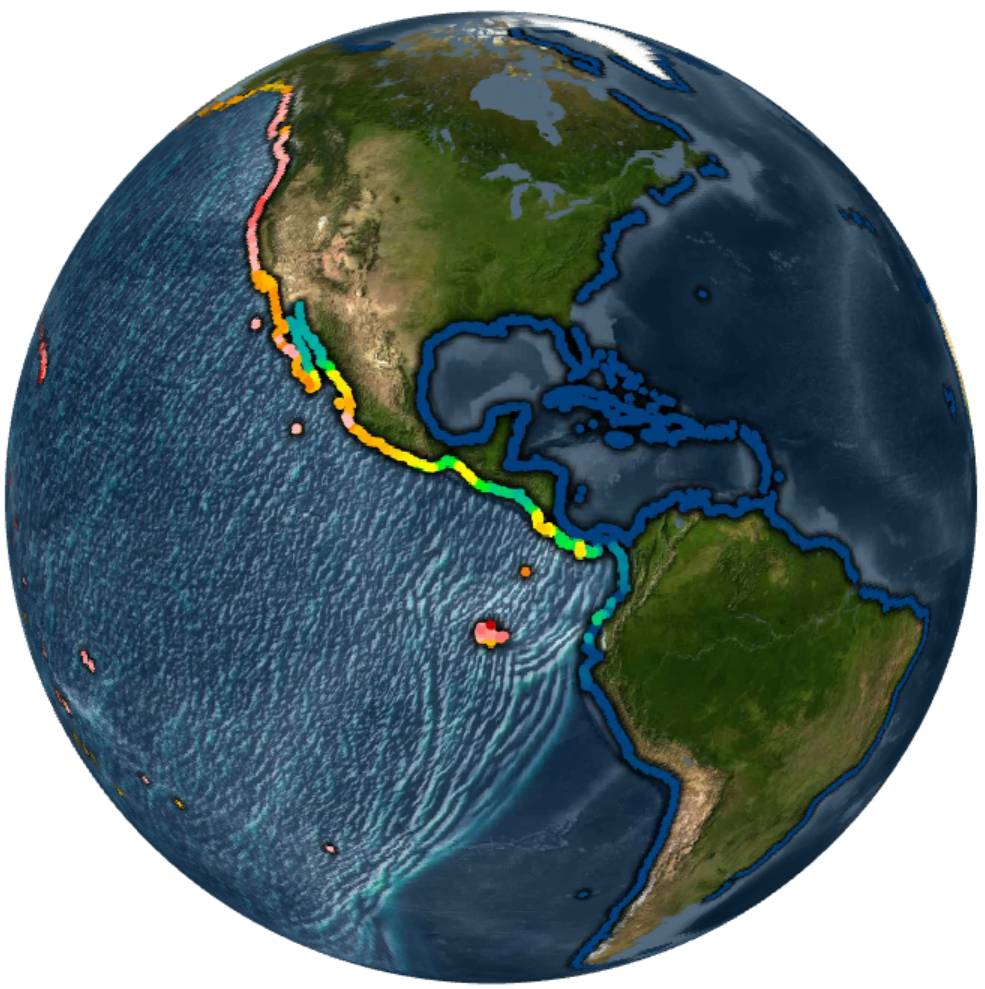}
    \includegraphics[width=\framewidth]{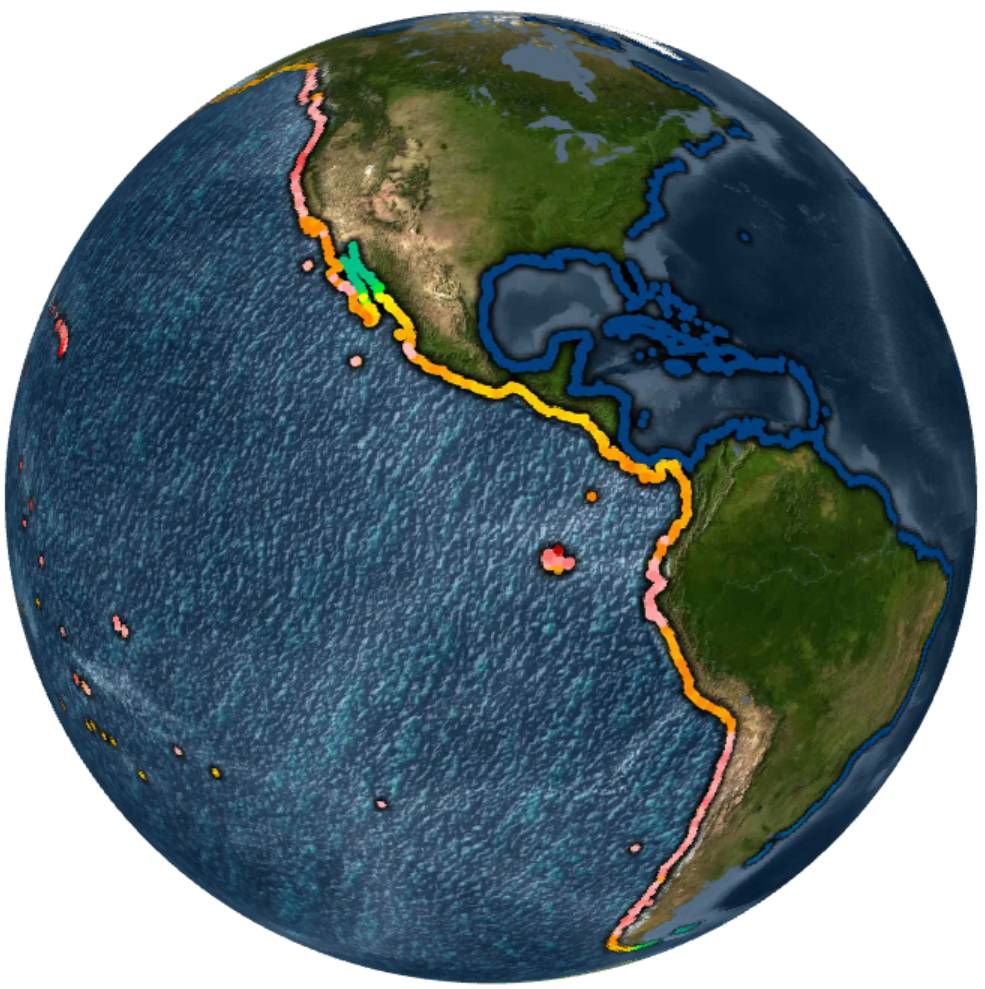}
    \includegraphics[width=\framewidth]{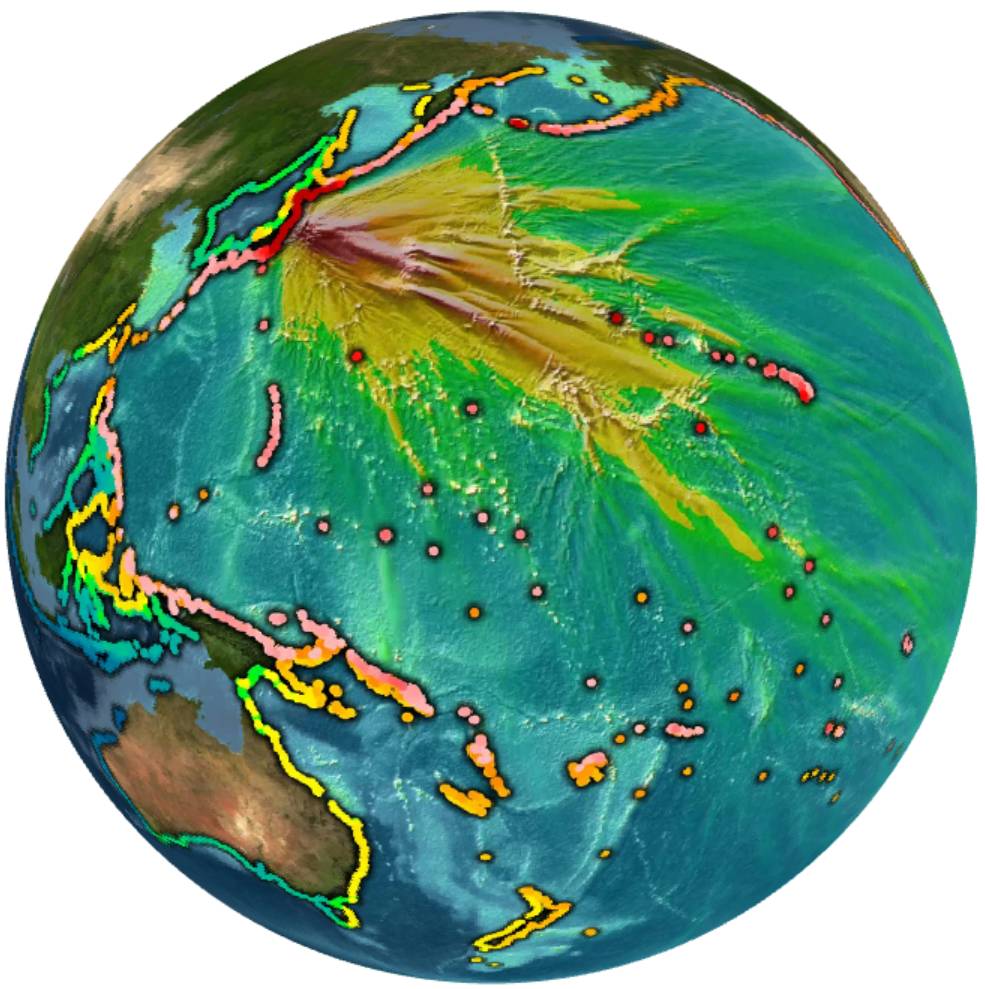}
    \includegraphics[width=\framewidth]{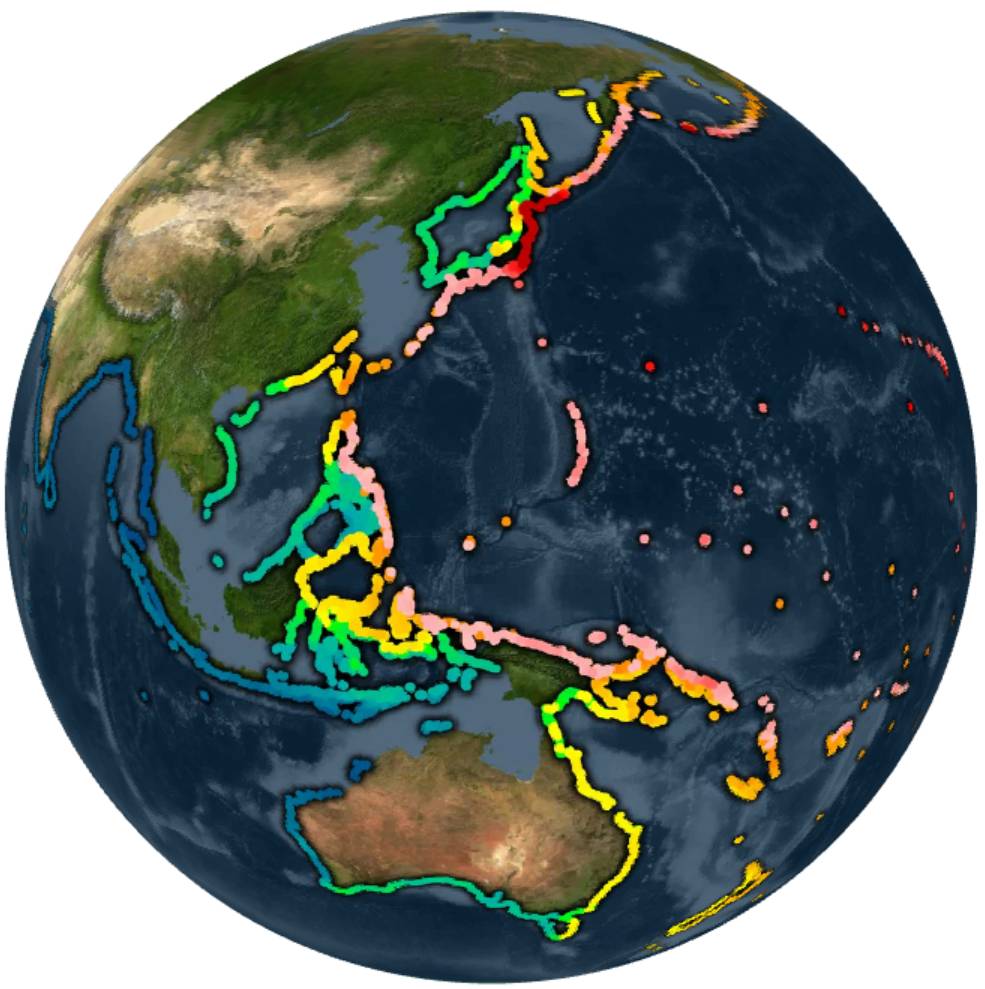}
 
    \caption{This \SoS dataset shows the spread of a tsunami across the Pacific Ocean, and its effects on various coastlines. Since the tsunami starts from a single point before expanding over the entire ocean, the visualization changes significantly over time, so the vision model requires more samples than the usual value of two. Without enough samples, essential information present at different time points could be missed, impacting the model’s ability to capture the full range of variations.}
    \label{fig:tsunami_frames}
\end{figure}

\subsubsection{Visual Information Extraction}
Once the frames are sampled, we feed them to the vision model with a suitable prompt, which instructs the model to extract information from the frames. We found that this model performs better when given detailed and context-specific prompts. Bendeck \etal~\cite{bendeck2024empirical} similarly observed that the performance of OpenAI's VISION model improves with structured, task-specific inputs, particularly when handling multi-step tasks like trend identification or comparison, perhaps because the specificity helps the model identify what is expected of it. However, manually fine-tuning prompts for each dataset would be too inefficient to satisfy \textbf{DR3—Extensibility}.

Therefore, we use the following prompt for all datasets: ``These are sample frames from the \SoS dataset. Identify all the possible visual cues, color patterns, and encodings if present. Then, generate a detailed description of the frame.'' This prompt makes the model generate descriptions of the visualizations, which generally correctly identify color and spatial patterns, such as marked features around or close to certain geographical zones, but without any indication of their meaning. This extracted information acts as a dictionary, establishing a direct relationship between the visual features and their corresponding values, thereby reducing the probability of incorrect responses. This process is illustrated on \cref{fig:json-img-relation}, and explained in detail in \cref{sec:json_description}.

\subsubsection{Merging and Structuring}
Once the visual information has been extracted, we append it to the SoS dataset description,\footnotemark{} and save the concatenated result. The original dataset description mainly covers the purpose behind the dataset, where and how the data were collected, and it may go deeper into the dataset's subject. However, it may not include all necessary visual encodings, so the extracted visual information helps by complementing the dataset description.

Once we have the concatenated information, we send it to the LLM within a prompt instructing it to generate a JSON file that includes all the relevant information and visual cues, and comprehensively describes the dataset. Here is the exact prompt: ``Given the information, generate a JSON file that contains all the relevant information, visual cues, and color encoding. Be descriptive enough that it can be understandable on its own. Include at least the following categories: Title, Description, Tags, Period of Time, Locations, Visual Cues, Color Encoding, Key Points, and Sources''. The novelty of this approach lies in the use of a structured format to insure that we get a single text file used for LLM augmentation. It combines the essential elements of the dataset's description with the visual information from the sampled frames, as shown on \cref{fig:json-img-relation}. This file is ready to be used by the runtime loop.


\subsubsection{Description of Example Augmentation File}
\label{sec:json_description}
\begin{figure*}[ht]     
    \centering
    \includegraphics[width=\textwidth]{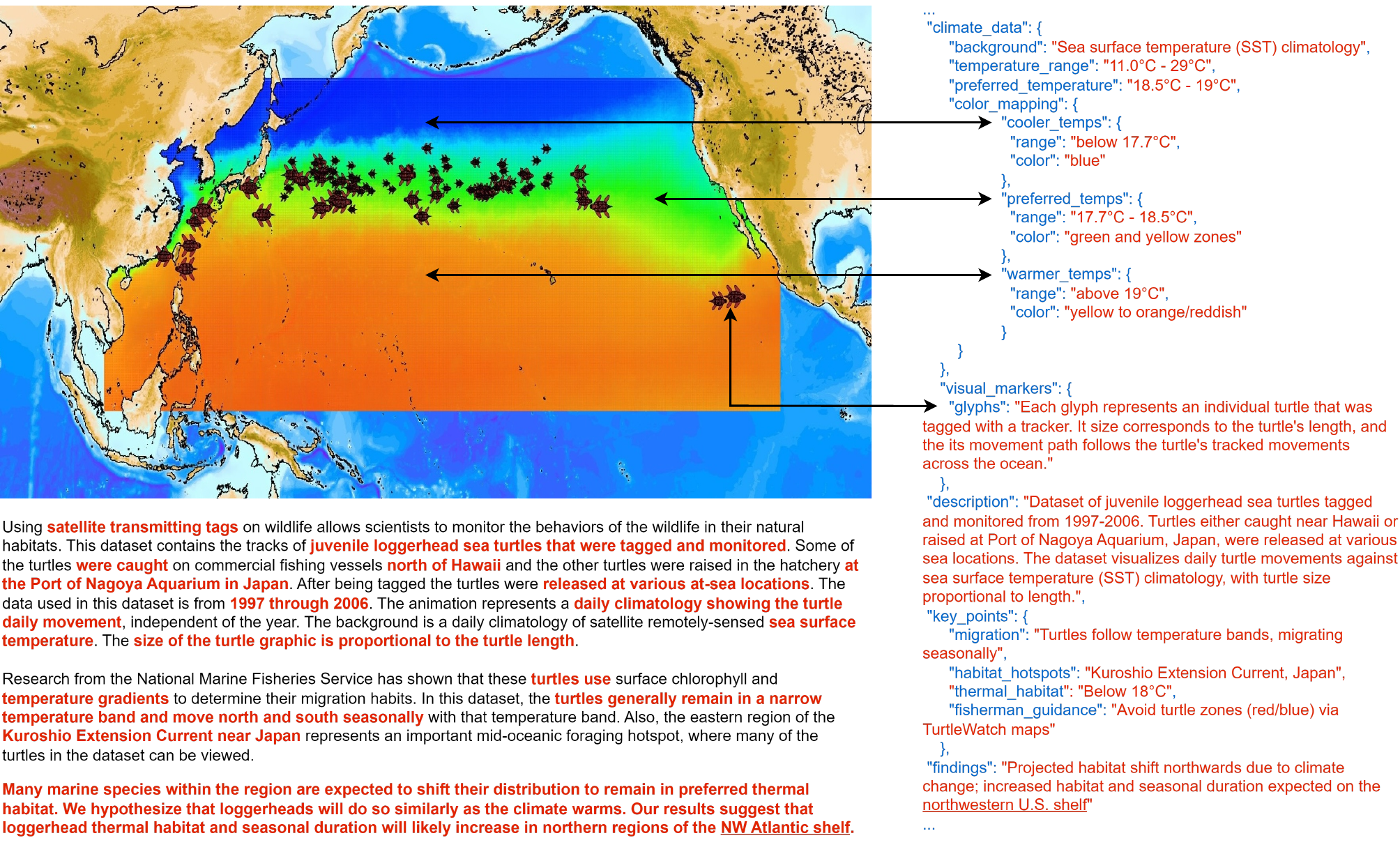}
    \caption{Example of augmentation data generation from the text description of a dataset and one of its rendered frames. This is the \myemph{Loggerhead Sea Turtle}. {\textbf{Top-left:}} A visual representation of sea turtle movements overlaid on sea surface temperature (SST) climatology data, with colors indicating temperature ranges and glyphs representing individual tracked turtles. The size of each glyph corresponds to the turtle's relative length. \textbf{Bottom-left} The text used to describe the dataset on NOAA's website, along with a subset of further relevant NOAA data, and of an associated publication~\cite{patel2021projected}. {\textbf{Right:}} A subset of the corresponding JSON description that encodes and summarizes both the visual and text input. It details climate data (SST, temperature ranges, and color mappings) and turtle tracking information (glyph sizes), and encodes key points of the text input. The parts of the input text that were included in the final JSON are highlighted in bold and red. The part that is also underlined was incorrectly synthesized by our method: GPT-4o erroneously noted the expected increase in the ``seasonal duration'' to be on the northwestern U.S. shelf instead of the Atlantic shelf. This may be because the dataset pertains to the Pacific, while this particular section of input text comes from the scientific publication, which deals with the Atlantic. Nevertheless, this illustrates the potential of structured data for LLM augmentation in a visualization context.}
    \label{fig:json-img-relation}
\end{figure*}

Here, to illustrate how our information extraction process works, and why the information it extracts is sufficient for augmentation, we provide a partial but detailed example of this process's output for a specific dataset, shown on \cref{fig:json-img-relation}.

The augmentation file (in JSON format) is a structured representation of one \SoS dataset, in this example, the Loggerhead Sea Turtles Tracking dataset. The file begins with basic metadata, including title, ID, and description (omitted on the figure). It gives a quick overview of the dataset's contents (turtle tracking data from 1997 to 2006), and its core focus on visualizing turtle movements against sea surface temperature (SST). A list of tags (also omitted) categorizes the dataset, facilitating searching and filtering based on relevant topics like ``marine biology'' and ``satellite tagging''. Additionally, the structure specifies the animal species and the time frame of their tracking study. It also specifies where the turtles were released and tracked, and provides background on sea surface temperature, detailing the temperature range mentioned in the dataset. One of the essential features of this structure is the color mapping, which links colors to numeric ranges.

The structure highlights migration patterns and habitat spots, noting that turtles follow temperature bands and migrate seasonally, with significant areas of interest like the Kuroshio Extension Current. The ``conservation\_status'' field (also omitted) notes the endangered status of loggerhead sea turtles under the ESA (Endangered Species Act), and the ``findings'' section discusses findings from a study conducted from 2009 to 2018, which indicates potential northward habitat shifts due to climate change, though the mention of increased habitat on the northwestern U.S. shelf is a mistake by the LLM, and should refer to the northwestern Atlantic shelf, based on input from a related article~\cite{patel2021projected}. The file concludes with a ``source'' field (omitted), attributing the dataset to the National Oceanic and Atmospheric Administration and the National Marine Fisheries Service, ensuring traceability and credibility.

The combination of OpenAI's VISION model and GPT-4 organizes the file with clear, separate sections for different types of information, allowing users and the system to quickly locate specific details about visual elements, color encodings, glyph meanings, etc., in the form of data representations intelligible to an LLM. This structure effectively integrates all relevant data into a coherent format, making it easy to reference and visualize complex information. This adds structure and clarity to our prompts to the LLM, but also makes them more compact (and therefore cheaper).

\subsection{Runtime: Conversational Interaction}
This runtime conversational interaction loop runs continuously while users interact with our system, receiving queries from users, and generating augmented prompts from these queries and the metadata extracted during the pre-processing step. First, the user's query is added to a context window containing a history of previous queries. This history is concatenated with the augmentation file generated in the pre-processing phase, and fed to two different LLMs, each tasked with a different assignment, and therefore provided with a different prompt: GPT-4o supplies the user with information, while GPT-4 converts user requests into commands to control the visualization, when requested. In response to the user’s query, the system always provides an informative reply to simulate conversation, with the option to include synthesized audio. Our proof of concept also features a conversational interface, so that depending on the nature of the query, the system may react, for example by rotating the globe to display a different region.

\subsubsection{Context Window}
Since our system relies on GPT-4o API calls for conversational interaction, we have no inherent memory capabilities like those found in the ChatGPT application: when it receives a new API prompt, GPT-4o is unaware of any previous ones. To circumvent this limitation and allow context retention over multi-turn dialogue, we added a window context system that creates a synthetic ``memory'' by including the last 20 interactions between the user and the LLM in every prompt, updating it dynamically as further prompts are issued --- we found this value of 20 to be sufficient for multi-turn dialogue, and low enough to avoid overwhelming the LLM with excessive information or confusing it regarding which instructions to follow.

This synthetic memory stores interactions between the user and the LLM in a FIFO queue. Each new interaction is pushed to the queue and embedded into every new request to maintain the context's history over time. Once this limit is reached, the oldest interaction is deleted, freeing up a new slot. This dynamic update of the memory over time is called a ``context window''~\cite{brown2020language}.


\subsubsection{Dual-Bot System}
In this system, every user query is relayed to two instances of GPT, each with a different purpose. The first instance (GPT-4o) processes the user's query to provide information. It acts as the conversational layer to deliver a clear, informative response. This is our main focus. The second instance (GPT-4) is responsible for transforming the user's query into a formal command that the system can deterministically parse and execute. As Jia \etal~\cite{jia2023voice} did with their \myemph{VOICE} project, we use a navigation bot to perform specific actions, such as providing coordinates for a place, changing the point of view, or changing the pose of the globe. The bot outputs precise numerical values, like geographic coordinates or an exact rotation angle.

For example, the system will output the coordinates if a user asks for a location on the map, as latitude and longitude, and optionally altitude. Positional commands are thus simply $\left[ \phi, \lambda \right]$ or $\left[ \phi, \lambda, r \right]$, where $\phi$, $\lambda$, and $r$ are floating point numbers. If the user asks for a different viewing angle, the system will return the numerical value for that specific viewpoint, in the following format: $\left[axis, a \right]$, where $axis$ is a string that can be ``x'', ``y'', or ``z'', and $a$ is a floating point number. This tells the system how much to rotate, and around which axis. If the query does not require any system action, the second GPT-4 instance will output "null", signifying that no further changes are needed to the current scene or perspective.


This simple approach allows the system to handle various user inputs effectively and avoids unintended changes when incorrect or malformed values are returned. This structured yet straightforward method is sufficient for the use case, ensuring the system can consistently interpret and execute commands while maintaining a responsive and error-tolerant user experience. By splitting the responsibilities for answering questions and executing commands, we effectively manage both the information and action-based aspects of user interactions. Having one system per task greatly reduces the risk of the LLM getting confused and failing to deliver the expected output. Interestingly, we found GPT-4 to be better than GPT-4o at following precise instructions to generate formal commands, hence our choice to use it for this purpose.

\section{Proof of Concept Implementation}
For ease of access across multiple platforms, we designed our system as a web application\footnotemark{} suitable for various devices:
desktop computers, mobile devices (smartphones), XR headsets, or globe displays.
It is deployed as a web application and contains a WebGL\footnotemark{} canvas, a chatbox for user interaction by text or voice (using a microphone), and a button enabling an XR session if the hardware supports it.
\footnotetext{\footnotesize{\url{https://www.khronos.org/webgl}}}
Upon loading the web application, users are presented with a globe showing the Earth and borders between countries. A message invites them to choose an animated \SoS dataset. Once selected, it is loaded and projected onto the globe, allowing users to interact with it.

If users converse by typing alone, only OpenAI API calls are needed. If voice I/O is desired, two different options are supported: A) Using the Speech Recognition function from WebSpeech API,\footnote{\footnotesize{\url{https://developer.mozilla.org/en-US/docs/Web/API/Web_Speech_API}}} native to the WebAPI itself, to transcribe the user's audio into text. This method transcribes fast (latency is around a second) but can be inaccurate, it can mix up (near) homophones, and only supports one language at a time, which must be configured before use. B) Using the MediaRecorder element, also native to WebAPI: it can only access the audio file once the user has finished speaking instead of transcribing it live, which we do with the Whisper\footnote{\footnotesize{\url{https://openai.com/index/whisper}}} API from OpenAI. This method is more accurate and works in all the languages the API supports, but is considerably slower: a query of 15 words can take up to 4 seconds to process, as we have to rely on remote processing from OpenAI servers running the Whisper Automatic Speech Recognition (ASR)\footnote{\footnotesize{\url{https://platform.openai.com/docs/guides/speech-to-text}}} model. 

Both methods use Text-to-Speech services from OpenAI to generate audio responses from LLM outputs. Rather than waiting for the whole file, the application receives streams of small audio chunks and plays them with minimal latency. To evaluate this reduced latency, we conducted fifty audio requests, measuring the time from the start of the request to the receipt of the first audio chunk and the start of the request to the entire audio generation. In the first case, we measured an average of 1.862 seconds, whereas for the second, an average of 4.090 seconds. This shorter latency makes the interactions feel more natural.

We chose JavaScript and its Parcel\footnote{\footnotesize{\url{https://parceljs.org}}} bundler, a web development tool that allows streamed media content (video frames) instead of requiring users to fully download all files before reading them. This helps us maintain a modular design divided into rendering, provided by WebGL, the synthetic memory, and audio processing, all of which share the same programming language.

For the back-end development, we chose FastAPI\footnote{\footnotesize{\url{https://github.com/fastapi/fastapi}}} because its developing environment makes it easy to create a server and test new endpoints. These endpoints bridge the JavaScript inputs with the LLM API requests.

Our application prototype uses WebGL to map the SoS geospatial animated textures on a sphere. We display an initial plain satellite image texture before any dataset is loaded.\footnote{\footnotesize{\url{https://www.shadedrelief.com/natural3/pages/textures.html}}}

We can map the sphere with any equirectangular map illustrating different aspects of the planet, such as CO$_{2}$ levels or average rainfall. The system can overlay additional images representing further information, such as timezones, continental or political borders, or capital names.



\begin{figure}[htb]
    \centering
    \includegraphics[width=0.66\columnwidth]{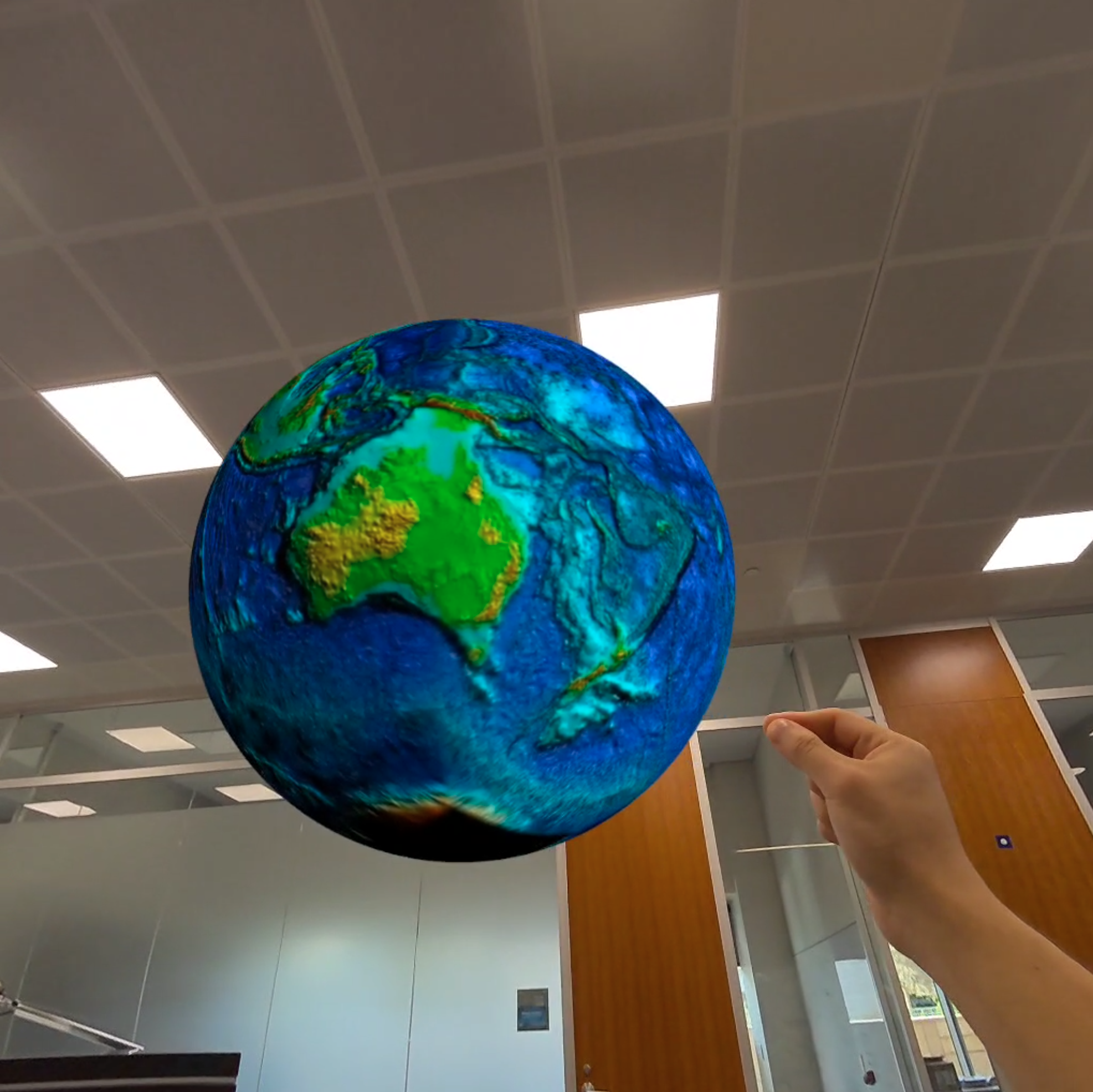}
    \includegraphics[width=0.33\columnwidth]{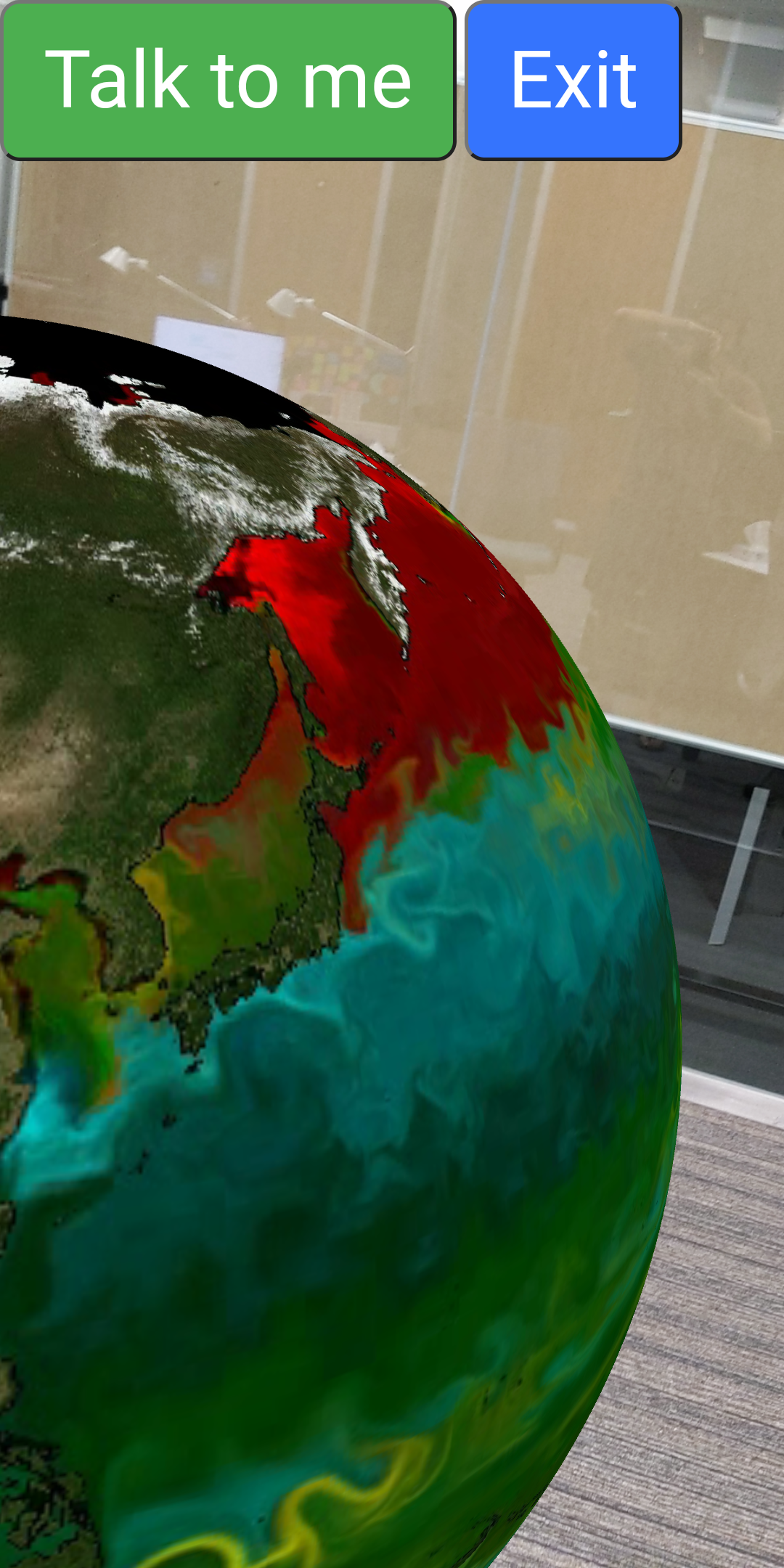}
    \includegraphics[width=0.66\columnwidth]{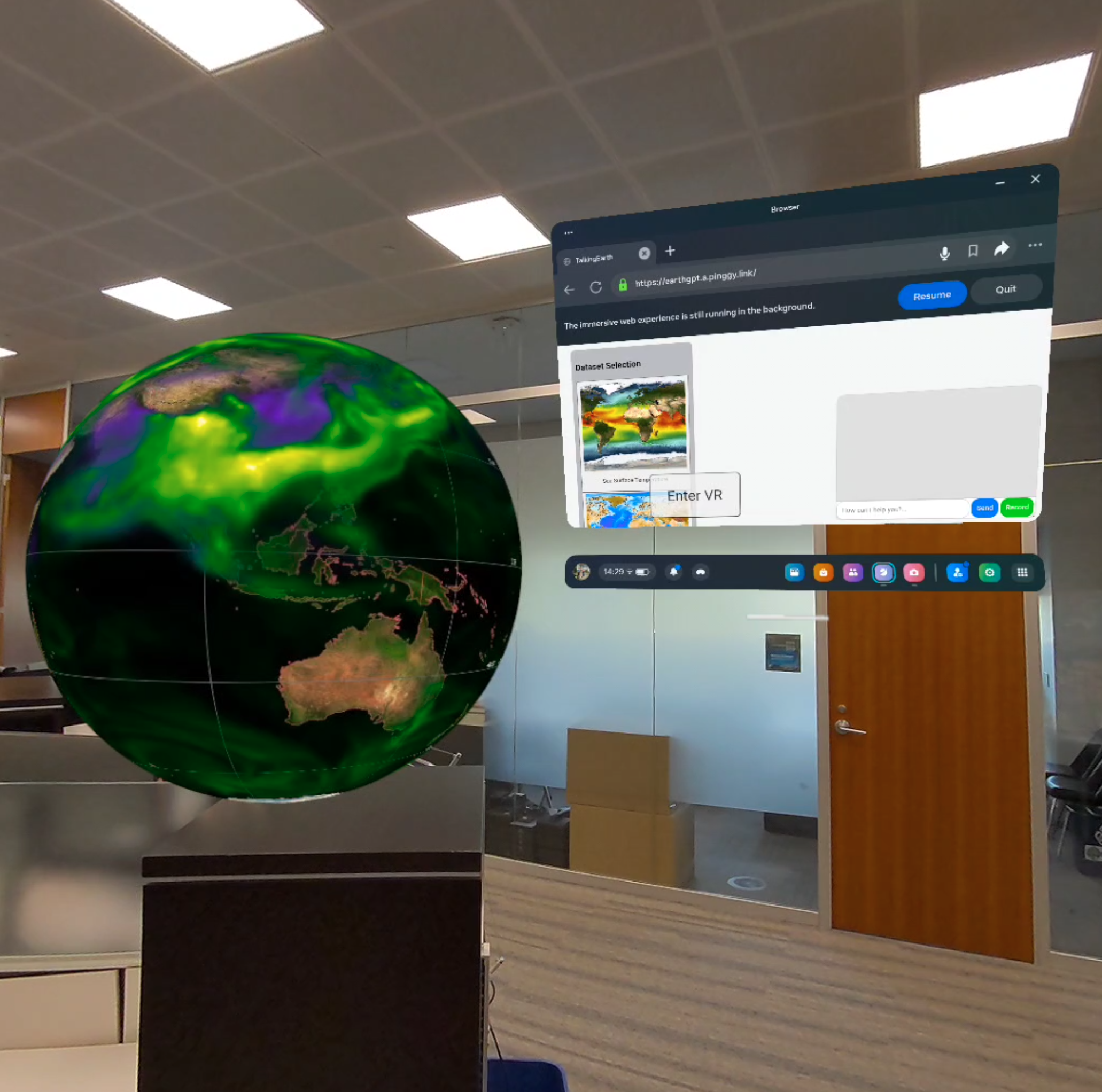}
    \includegraphics[width=0.33\columnwidth]{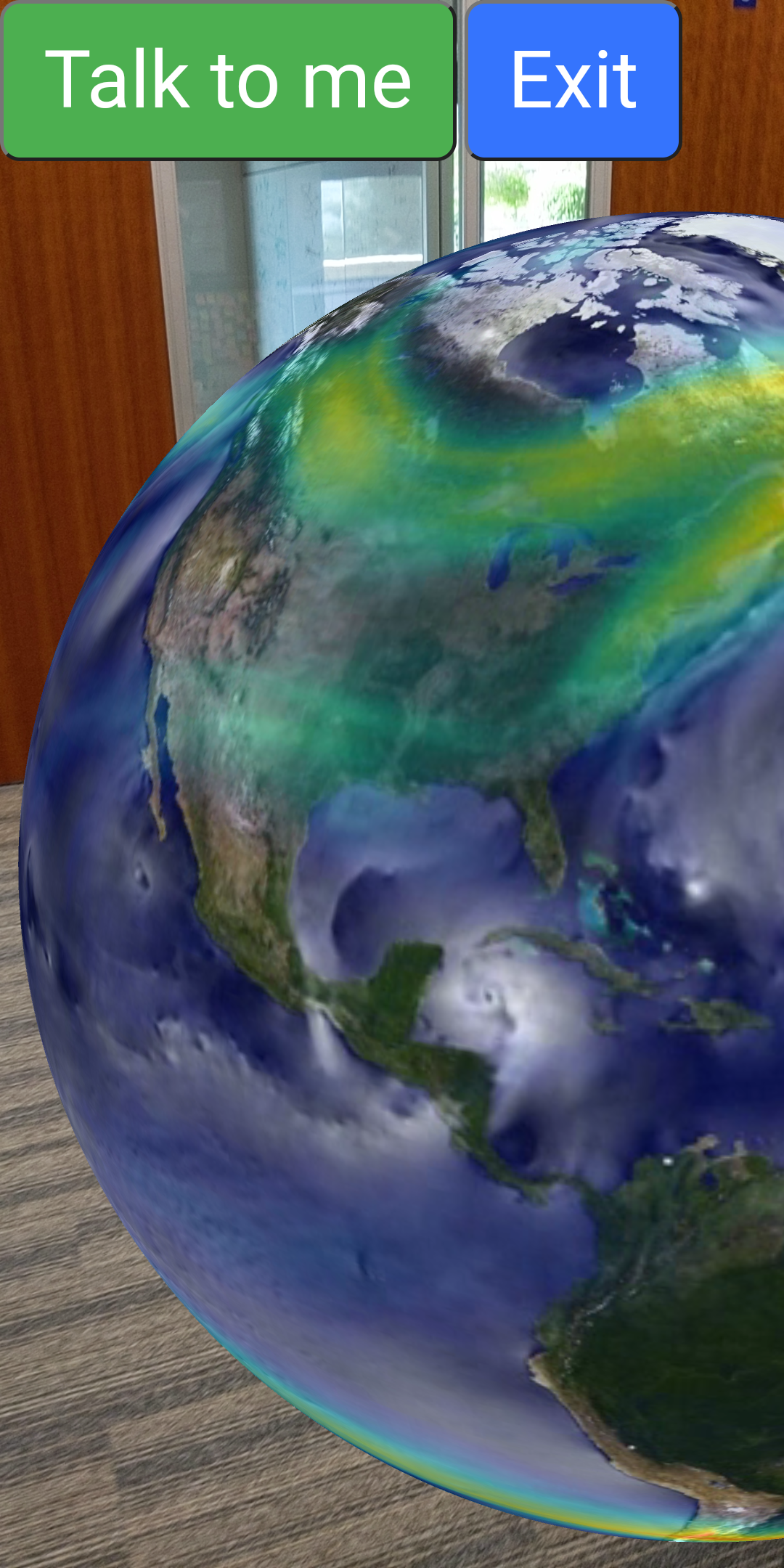}
    \caption{Our proof of concept showing geospatial visualizations on different devices. Meta Quest 3 is on the left, and Google Pixel is on the right, both in augmented reality. \textbf{Top-left} shows a topographic map. \textbf{Top-right} shows concentrations of phytoplankton. \textbf{Bottom-left} shows presence of black carbon in the atmosphere, along with the floating webpage app. \textbf{Bottom-right} shows world-wide wind patterns.}
    \label{fig:xr_examples}
\end{figure}

\newcommand{\topwidth}{0.23\textwidth}
\newcommand{\subwidth}{0.32\textwidth}

\begin{figure*}[htb]
    \centering
    
    \begin{minipage}{\textwidth}
        \centering
        \begin{subfigure}[b]{\topwidth}
            \centering
            \includegraphics[width=\textwidth]{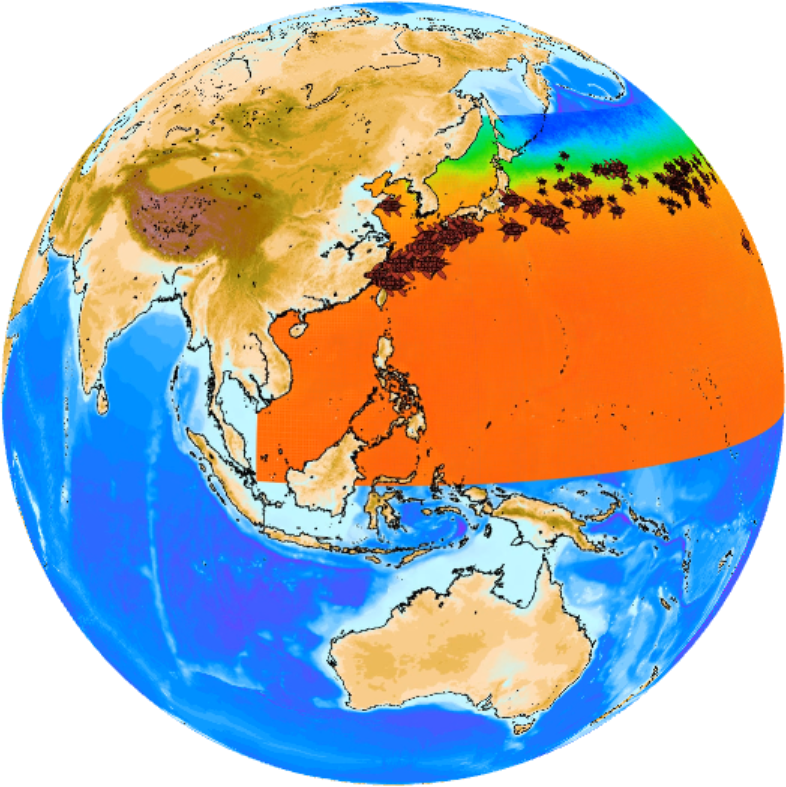}
        \end{subfigure}
        \hfill
        \begin{subfigure}[b]{\topwidth}
            \centering
            \includegraphics[width=\textwidth]{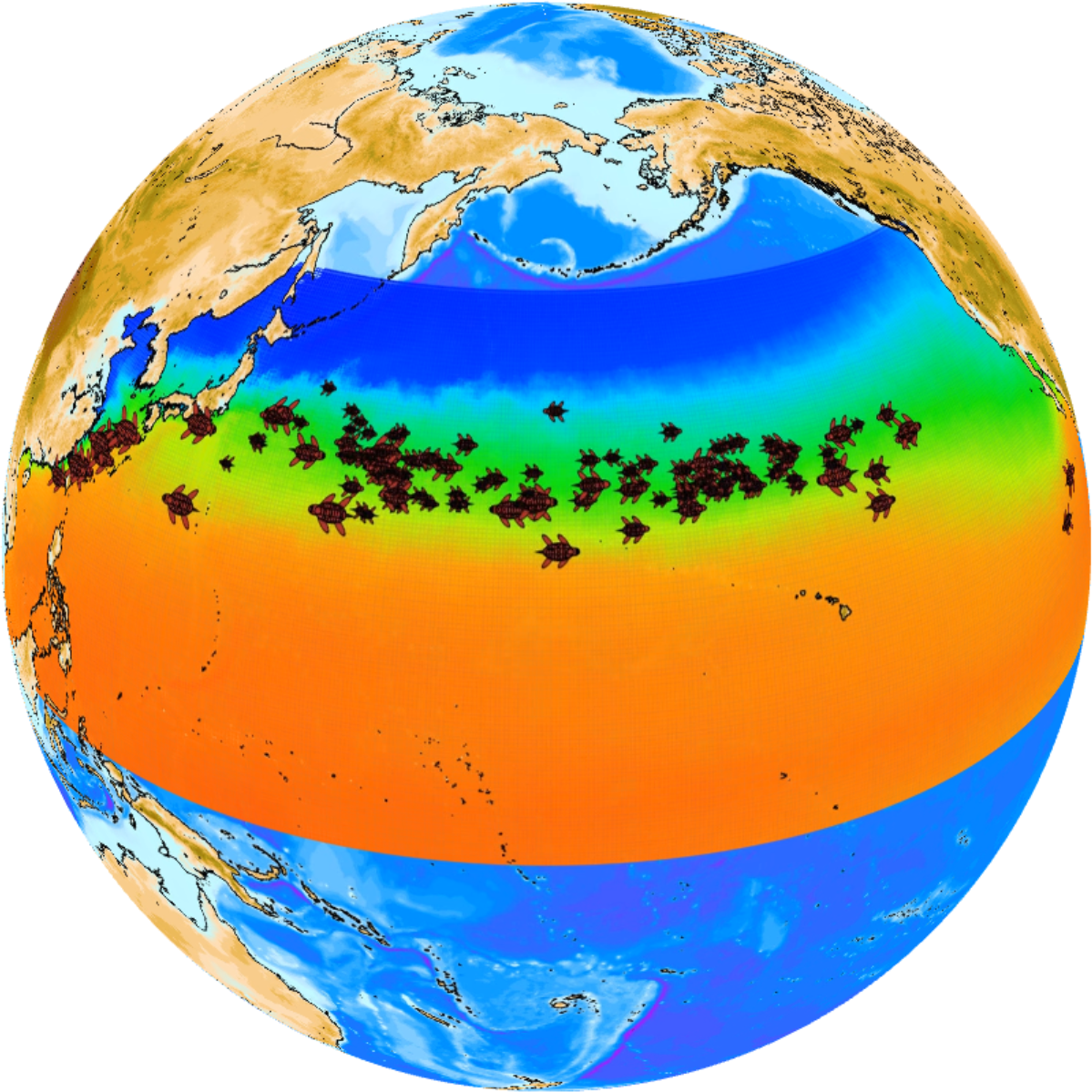}
        \end{subfigure}
        \hfill
        \begin{subfigure}[b]{\topwidth}
            \centering
            \includegraphics[width=\textwidth]{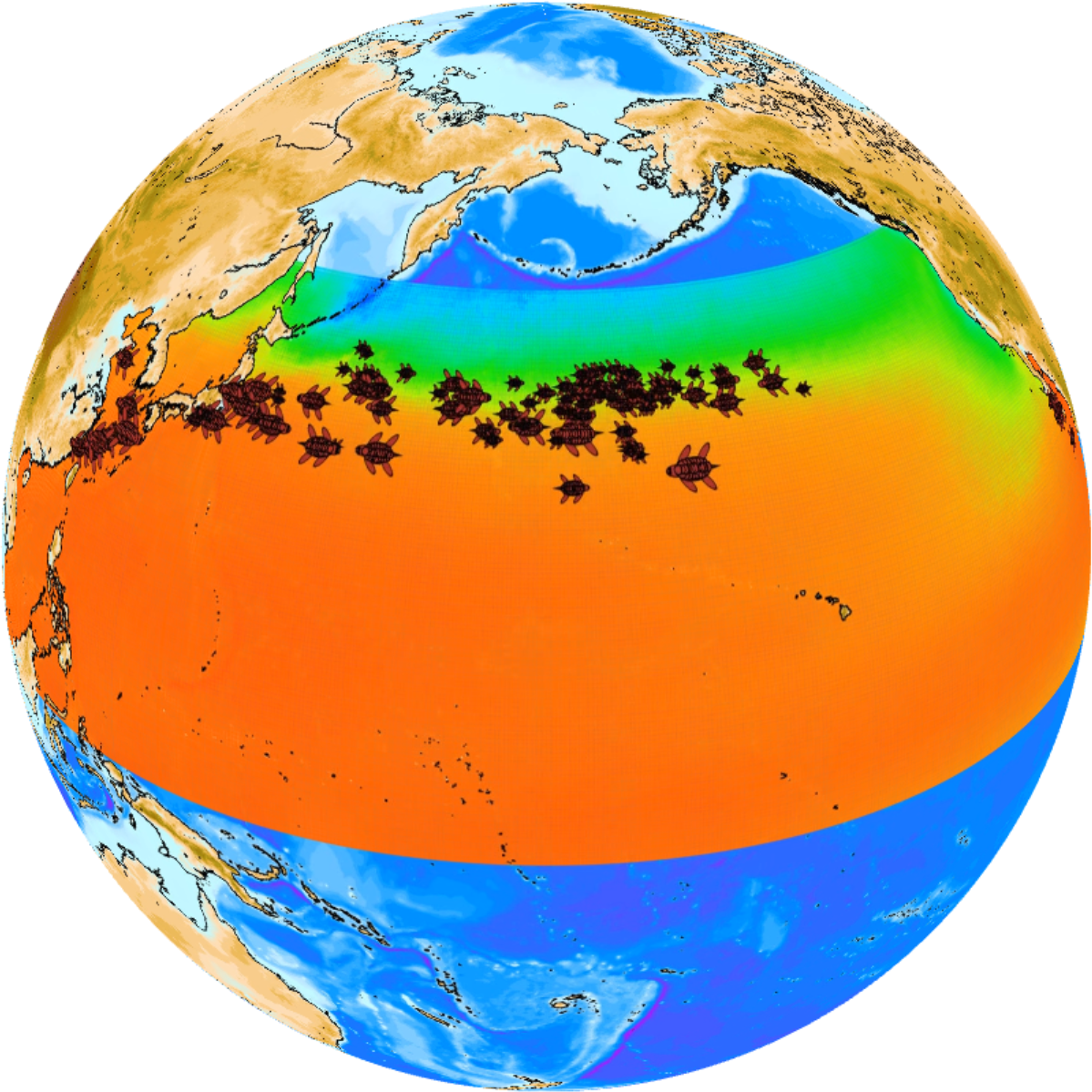}
        \end{subfigure}
        \hfill
        \begin{subfigure}[b]{\topwidth}
            \centering
            \includegraphics[width=\textwidth]{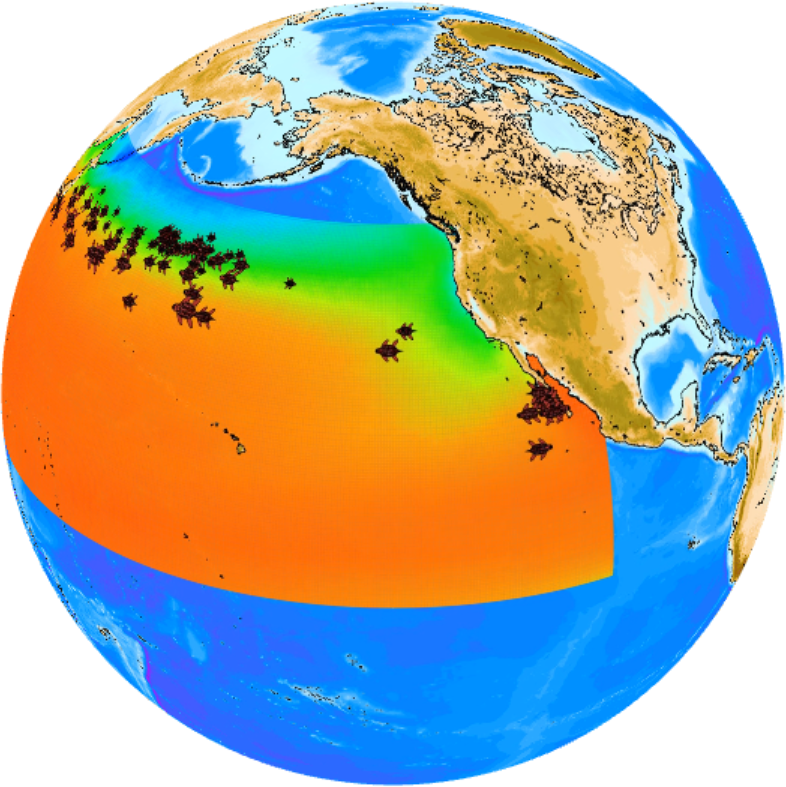}
        \end{subfigure}
        \label{fig:demo_vis}
        \caption{Visualization of the turtle migration dataset at different stages, and from different angles.}
    \end{minipage}

    \begin{subfigure}[b]{\subwidth}
         \centering
         \includegraphics[width=\textwidth]{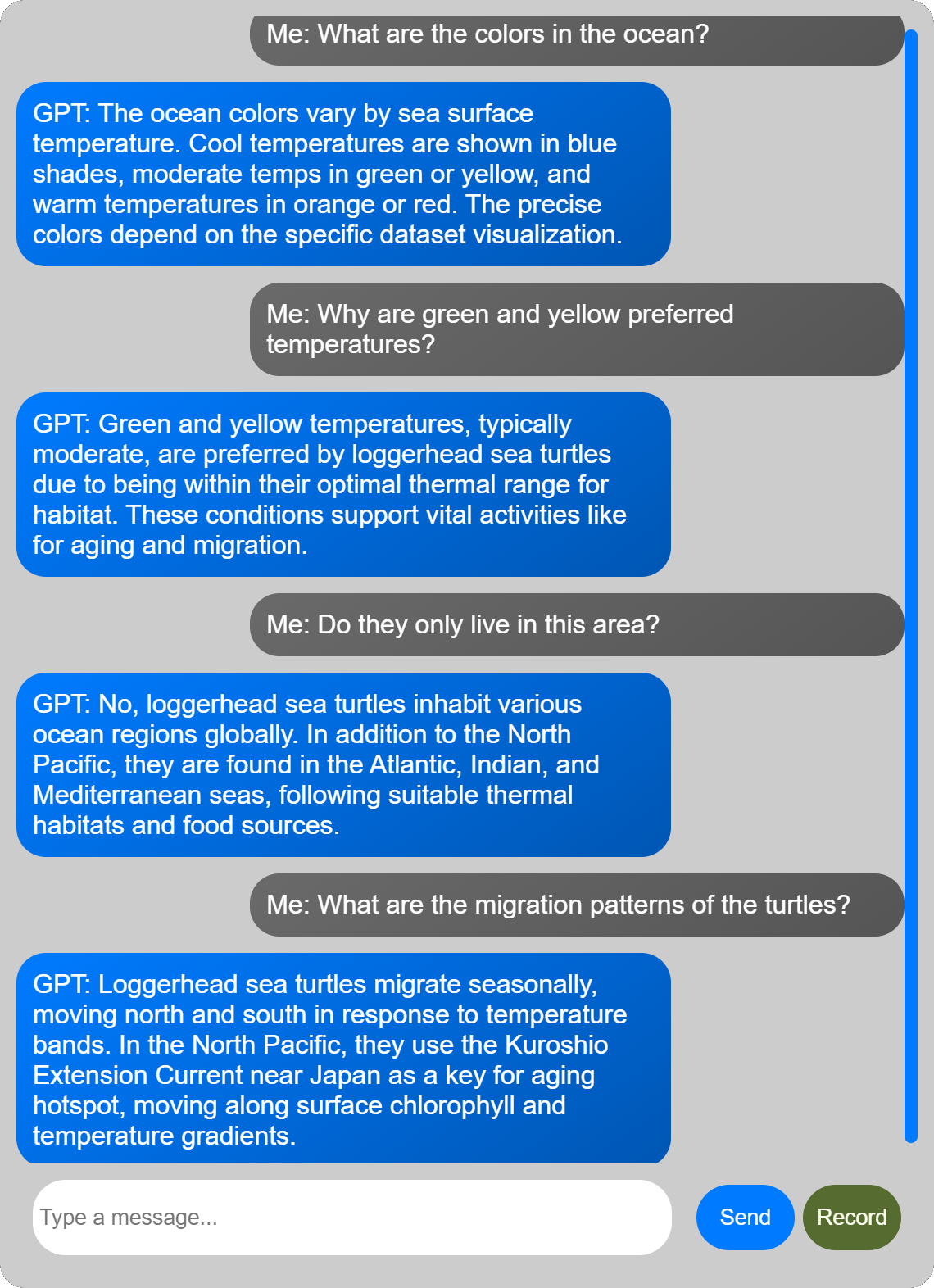}  
         \caption{Text augmentation only.}
         \label{fig:demo_TextOnly}
    \end{subfigure}
    \hfill
    \begin{subfigure}[b]{\subwidth}
         \centering
         \includegraphics[width=\textwidth]{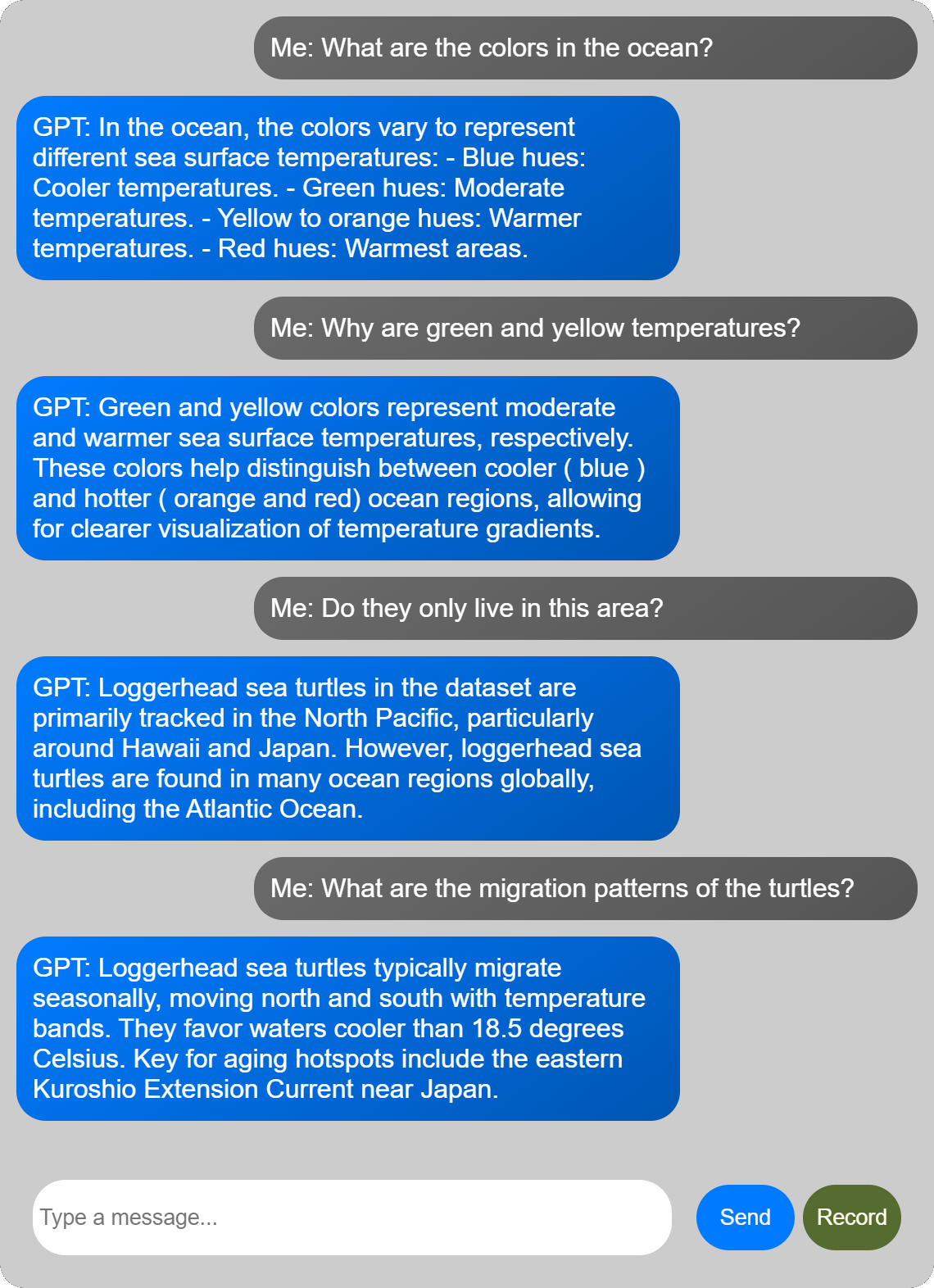}  
         \caption{Text + image augmentation.}
         \label{fig:demo_TextImage}
    \end{subfigure}
    \hfill
    \begin{subfigure}[b]{\subwidth}
         \centering
         \includegraphics[width=\textwidth]{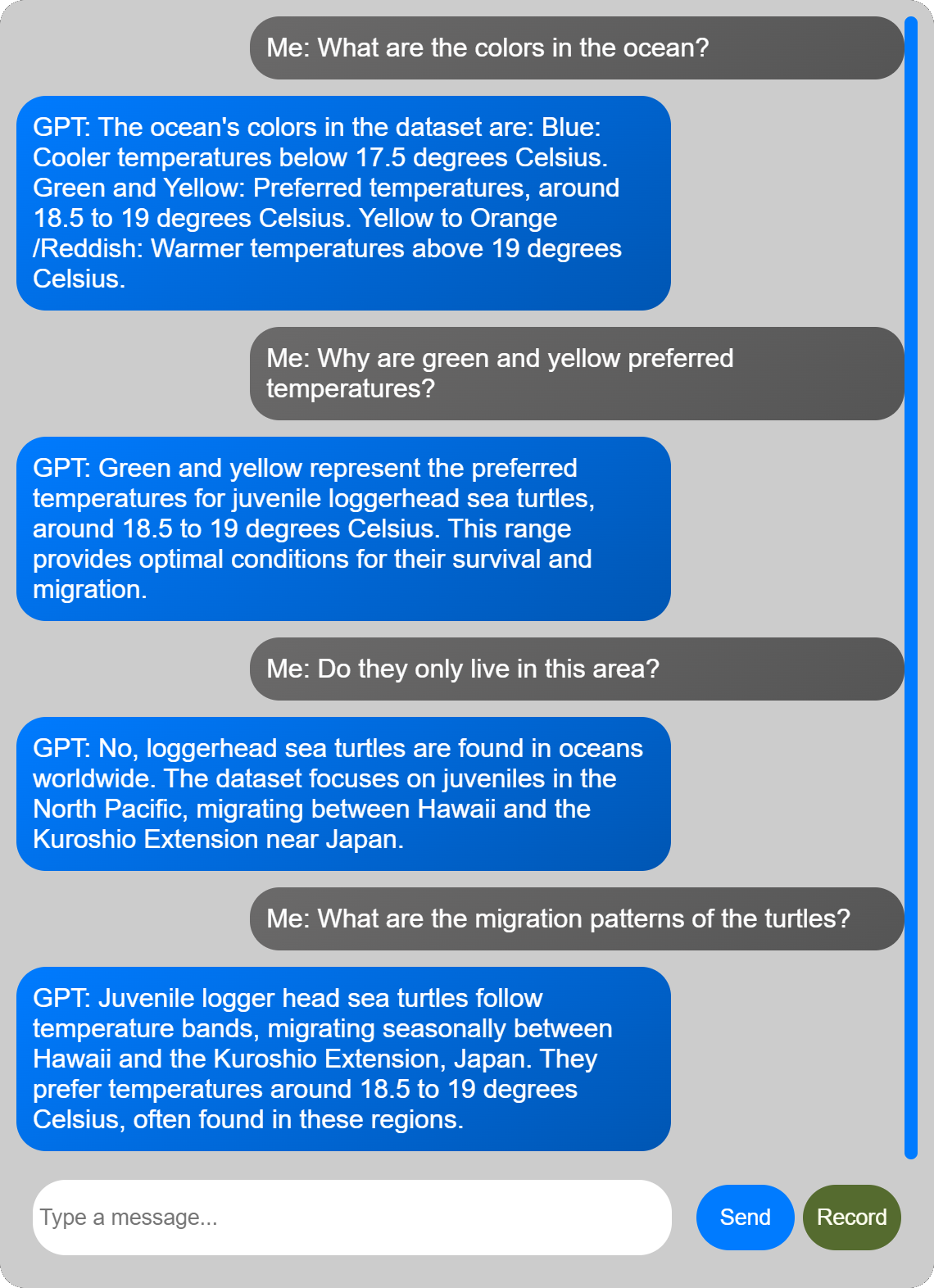}  
         \caption{Structured data augmentation.}
         \label{fig:demo_JSON}
    \end{subfigure}
     
    \caption{Demonstration of our system in action when the user is asking questions relating to the dataset about loggerhead sea turtles. This figure shows the different responses depending on the type of augmentation used, showing the benefits of our augmentation based on structured data incorporating information from both the description of the dataset and rendered snapshots.}
    \label{fig:demo_triple}
\end{figure*}

\section{Demonstration}

The proof of concept we developed to demonstrate our method supports three configurations: desktop, Augmented Reality (AR), and Virtual Reality (VR), depending on the device used. In AR and VR environments, users can interact with the system through voice commands and hand gestures, enabled by hand-tracking technology. For example, gestures like the pinch motion allow users to manipulate the Earth by translating or rotating it using wrist movements. These interactions are facilitated by retrieving hand position and orientation data from the WebXR session. \cref{fig:xr_examples} showcases our proof of concept (PoC) in an XR environment, highlighting physical interactions as users visualize ocean temperatures and hurricane patterns.

\subsection{Structured Augmentation: Sea Turtles}
To further demonstrate the method, \cref{fig:demo_vis} and \cref{fig:demo_triple} provide an example of a conversation with the system while visualizing a loggerhead sea turtle migration pattern dataset. The conversation is based on a real interaction from our user study, where a participant explored the dataset using voice and hand input. During the session, GPT-4o provided information about the visualized data. Without our structured data augmentation, the model provided vague responses, such as imprecise temperature ranges for the color bands. Even when visual information was included, GPT-4o struggled to offer specific details. However with structured augmentation, our system correctly identified the link between the green/yellow bands and the $18.5$ to $19$ \degree{}C range, which is the preferred water temperature for the turtles.

This specific \SoS dataset features a rather succinct and incomplete description\footnote{\footnotesize{\url{https://sos.noaa.gov/catalog/datasets/loggerhead-sea-turtle-tracks}}}. This is unusual and it may be because it was among the earliest datasets included in the SoS project. So, we supplemented it with additional information (whether structured augmentation was used or not). The ``Data Source'' section on this site directs us to the NOAA Fisheries website\footnote{\footnotesize{\url{https://www.fisheries.noaa.gov/about/pacific-islands-fisheries-science-center}}}, but information about loggerhead turtles is dispersed across the website. We looked for the specific species in the dataset\footnote{\footnotesize{\url{https://www.fisheries.noaa.gov/species/loggerhead-turtle}}}. The website contains loggerhead sea turtles information pertaining to both the Pacific and Atlantic Oceans. Pursuing comprehensive details on their migration patterns and shifts in a thermal habitat, which are species-specific, we found relevant information mentioned by Patel \etal~\cite{shiftsTurtles}. However, the information there is still scattered, so we explored further until we found the website that contains the explicit visual encodings \footnote{\footnotesize{\url{https://oceanwatch.pifsc.noaa.gov/turtlewatch.html}}}. Therefore, we need to manually verify if the information on the \SoS webpage\footnote{\footnotesize{\url{https://sos.noaa.gov/catalog/datasets}}} is sufficient or if we should supplement it with additional details from official sources.

When the LLM was only augmented with text and visuals, it incorrectly stated that turtles prefer waters cooler than $18.5$ \degree{}C. Our system, however, accurately provided the correct information, specifically mentioning the $18.5$ to $19$ \degree{}C range. This example highlights how our method enhances the accuracy of the LLM's responses, particularly in the context of scientific datasets, ensuring more reliable and informative interactions for the user.

This interaction takes place with fairly low latency, with complete text responses reaching the user in under a second. With speech synthesis enabled, audio messages start streaming in within a couple of seconds after submitting a query to the system.

\subsection{Highly Varying Data: Tsunami in the Pacific}
As discussed in \cref{sec:sampling}, some SoS datasets exhibit very significant changes over time, and this is the case of the one shown on \cref{fig:tsunami_frames}. It represents the spread of the tsunami caused by an earthquake near Japan, in 2011. It contains data showing the wave moving across the ocean over some 41 hours, and the coastlines that were struck, with red and yellow indicating the greatest impacts, particularly in Japan, where some of the waves exceeded 40 meters in height. For such datasets, sampling more than two frames from the rendered visualization is preferable. In this case, the wave moves from a single point to the entire ocean, so the LLM will not be able to provide information about the different stages of the dataset without sampling multiple frames, and making sure they are spread over the 41 hours recorded in the dataset.

\subsection{Personal Computers and XR Devices}
Our proof of concept was designed to run on multiple platforms from the ground up, and supports personal  virtual and augmented reality, both on dedicated devices and on capable smartphones. This is shown on \cref{fig:xr_examples}. On the left, our PoC is running on a personal computer, displaying the locations of the temperature measurements that were used in the 2023 Goddard Institute for Space Studies Surface Temperature Analysis v4, showing the different types of devices used and their associated colors. The middle image is a screen capture from a smartphone displaying ``how the global ocean's surface water temperatures vary over the course of few years, [\ldots{} and] how surface ocean currents and eddies transport heat and water around the globe. The images were generated [\ldots{}] from [\ldots{}] computer model of Earth's climate created at NOAA''. The rightmost portion of the figure demonstrates a Meta Quest 3 device displaying a visualization of water vapor, as hourly time steps, based on ``a fully coupled, global climate model that provides state-of-the-art computer simulations of the Earth's past, present, and future climate states''.

Figure~\ref{fig:xr_examples} further illustrates the capabilities of our system. On the top-left, the visualization rendered on a Meta Quest 3 ``gradually reveals the sea floor as the ocean is `drained.' The scale in the dataset shows the distance below sea level in meters and miles.'' On the top-right, a smartphone running our system shows ``dominant phytoplankton types from 1994--1998 generated by [\ldots{}] a high-resolution ocean and ecosystem model. [It] contains flow fields [\ldots{}], inorganic nutrients, 78 species of phytoplankton, zooplankton, as well as particulate and dissolved organic matter. Colors represent the most dominant type of phytoplankton at a given location based on their size and ability to uptake nutrients.'' Bottom-left is a visualization  where ``scientists have been analyzing the presence of black carbon in the atmosphere'' in 2007. At the bottom-right is a visualization of ``the speed of winds at the tropopause'' in 2006--2007. ``Such simulations allow scientists to view the intensity and turbulence of the polar and sub-tropic jet streams, which carry weather around the globe. Red, orange and yellow are used for the fastest moving air.''

Our proof of concept also supports virtual reality, but on a screenshot, there is no discernible difference between this setup and a standard personal computer.

\section{Hypotheses and Evaluation}

The process we developed to extract essential features and visual information from a dataset's description is meant to augment an LLM with a combination of text and visual data, so that it will produce accurate answers. We thus formulate the following hypotheses:

\textbf{H1:} Our proposed method will prove significantly more accurate than GPT-4o augmented with a dataset's full description, but without visual information, when asked to describe a particular visualization and explain how to read and interpret it.

\textbf{H2:} Our proposed method will prove more accurate than GPT-4o augmented with a dataset's full description \myemph{and} visual information when asked to describe a particular visualization and explain how to read and interpret it.

\textbf{H3:} Our PoC will be well liked by the participants, and receive a SUS score above the average of 70 measured by Bangor \etal~\cite{bangor2008empirical} over more than 2000 surveys.

\textbf{H4:} Specifically, participants will deem our application \myemph{not} to require the support of a technical person for its use, that learning how to use it would be very quick, and that they do not need to learn a lot of things before they could get going with it. These SUS items pertain to the key metrics for our system, given the general goals of the \myemph{TellUs} project.

\subsection{Evaluation and Experimental Protocol}

Our method features augmentation for GPT-4o, especially visual augmentation, to ensure the LLM can provide complete and accurate information about each dataset. While augmenting LLMs with text information for visualization is becoming a common practice, visual augmentation is not. The primary purpose of our evaluation is to assess the effect of this visual augmentation on the quality of the LLM's responses, we submit real user prompts to both our proposed vision-augmented JSON-based prompt and to a baseline method where the full-text description of a given dataset is part of the prompt, either with or without the image, while our method has access to the JSON metadata file instead.

We recruited 17 participants (average age of 26.53, standard deviation of 3.33) and asked them to try our application. We obtained informed consent in writing from all participants. We used a laptop to conduct the study. During the sessions, we briefed the participants to ensure they would understand the application and know how to interact with it. We were present to offer technical support, but we did not limit the users in how they should interact. Instead, we encouraged them to think beyond the immediate dataset by asking broader and relational questions to the augmented LLM, though without suggesting specific topics or questions.

We collected all of the statements they addressed to the facilitator, as well as the facilitator's output, and submitted the exact same utterances in the same order to GPT-4o, after providing it with a description of the dataset, either with or without the image's features (extracted with OpenAI's VISION model). We collected its outputs, and graded all outputs for correctness, on a scale of 1 to 10. To reduce the risk of bias when evaluating the outputs, the experimenter tasked with doing this was blinded to the source of each output, only being asked to grade outputs A, B and C for each prompt, without being told which system had produced it. The association between A/B/C and a given system was picked randomly for every single prompt. The first author, who had spent a great deal of time testing the system, was not tasked with this evaluation, to minimize the risk of compromising the blinding process through familiarity with any particular style of output. The person grading the LLM outputs read the dataset's description and associated scientific publications thoroughly, and studied the topics of each prompt until they felt confident they could assess the correctness of each output.

Though the primary reason for recruiting participants was to gather a set of real user queries, and thus avoid the bias that might come from generating them ourselves, we seized this opportunity to get some qualitative feedback about our proof of concept. After each participant had spent 20 minutes interacting with the system, we conducted a short interview to gather their impressions, and asked them to answer a System Usability Scale~\cite{brooke1996sus} questionnaire.

\subsection{Evaluation Results}
\begin{figure}[ht]
    \centering
    \includegraphics[width=\columnwidth]{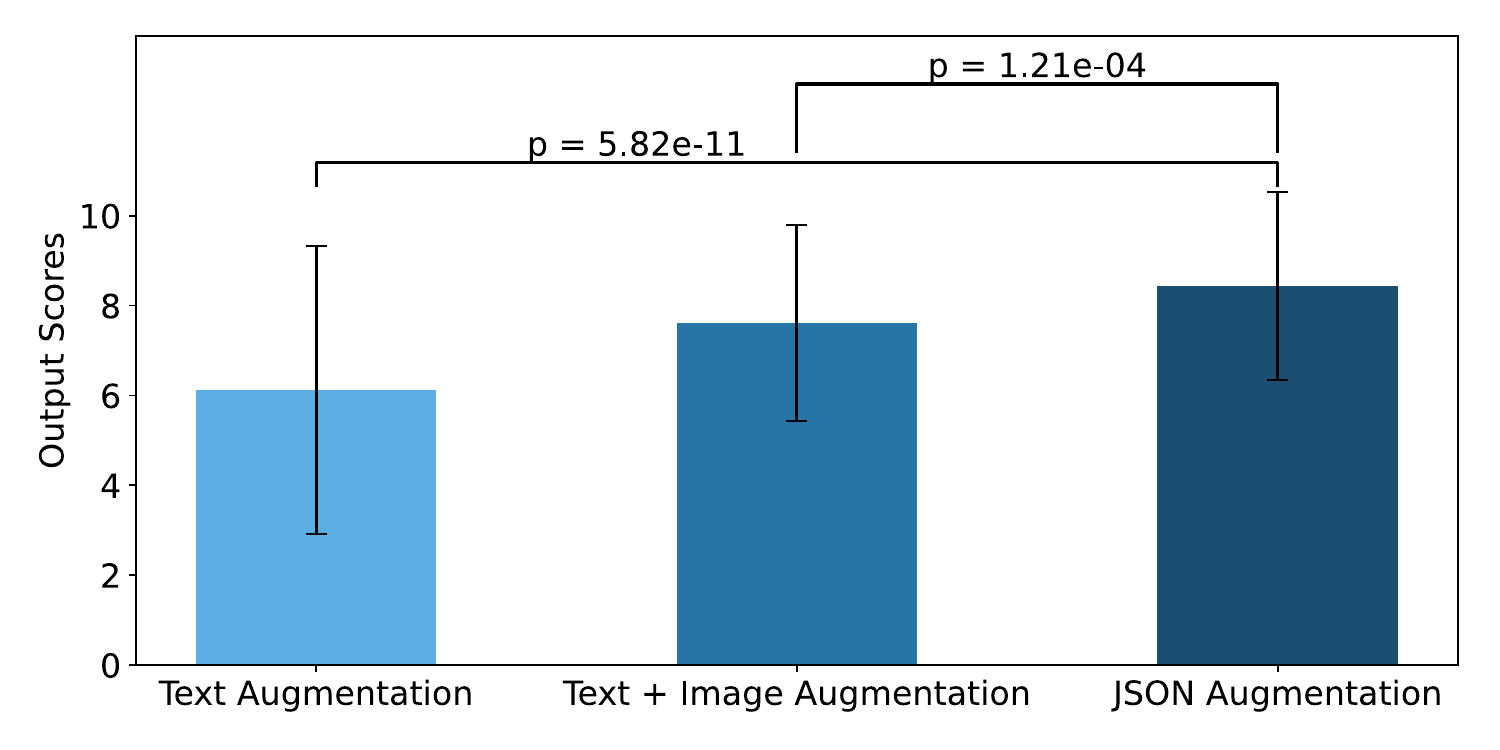}
    \caption{Evaluated accuracy of GPT-4o with three different types of augmentation, with standard deviations shown as error bars, and p-values as annotations above.}
    \label{fig:output_plot}
\end{figure}

This provided us with 125 user queries, and therefore one output per query and per condition. We evaluated the outputs produced by our PoC and both baselines, giving them each a score from 0 (worst) to 10 (best). Our data significantly deviated from normality, as indicated by Shapiro-Wilk test~\cite{shapiro1965analysis} p-values well below $0.00001$, so we used the Mann-Whitney U test~\cite{hollander2013nonparametric} to compare the distributions for each pair of conditions. We report the results on \cref{fig:output_plot}. We display the mean values and standard deviations, as well as p-values computed from the Mann-Whitney U test. After Bonferroni correction,~\cite{rupert2012simultaneous} these remain significant.

Our PoC is significantly more accurate than pure text augmentation, supporting \textbf{H1}. It also outperforms the baseline method that combines text description with visual information, even though the baseline includes more data overall, supporting \textbf{H2}. This highlights the importance of using a structured JSON format for effective augmentation. which is consistent with the insight gained during our development phase: it is not necessary to provide a lengthy and extremely detailed description for the LLM to interpret the prompt and provide an accurate explanation of the requested topic. Rather, the key seems to lie in adapting our prompts and interactions to align with the architecture and inherent strengths of LLMs. These language models do not process knowledge as humans do since their ``knowledge'' is derived from extensive datasets, and their architecture is based on statistical relationships~\cite{vaswani2017attention} within data.

By accounting for this difference, we can optimize our interactions with an LLM by not overwhelming the model with verbose explanations. We can obtain equally effective results by crafting our prompts concisely and systematically to align with how LLMs process information. This JSON-based approach allows us to extract and present only the most relevant information from a dataset, ensuring the LLM's responses are accurate and contextually appropriate. By doing so, we leverage its capabilities of prompt processing effectively, and perhaps more so than with plain text, even if said plain text may include more information. If our interpretation of this result is correct, it emphasizes the importance of understanding the specific nature of LLM knowledge processing and of adapting to it.

To gather qualitative insight about the overall experience of using our PoC, we collected SUS scores from our participants. The average SUS score measured across all participants is 83.75, with a standard deviation of 4.73, supporting \textbf{H3} as well.

Given the eventual goals of the \myemph{TellUs} project, we also considered the most relevant individual items from the SUS questionnaire, and we report the results here:
\begin{description}
    \item[4:] ``I think I would need the support of a technical person to be able to use this system'': mean score of 1.17, (1 means ``strongly disagree'') with a standard deviation (stddev) of 0.37.
    \item[7:] ``I would imagine that most people would learn to use this system very quickly'': mean score of 4.5, (5 means ``strongly agree''), stddev: 0.5.
    \item[10:] ``I needed to learn a lot of things before I could get going with this system'': mean score of 1.17, stddev: 0.37.
\end{description}
These results support \textbf{H4}, and are very encouraging for future deployment of virtual facilitators.

\section{Limitations \&{} Future Work}

Our proof of concept demonstrates a robust method for augmenting an LLM with a coherent combination of text and visual data, allowing it to perform the basic function of a human facilitator. However, compared to a real human facilitator, it remains limited in a number of important ways.

\subsection{Augmentation Method Limitations}
Some of those limitations pertain to the augmentation method itself. For instance, while our system is aware of the general appearance of a visualization, it does not have time-dependent visual information: it does not know what the visualization is showing at any specific time, only what it looks like in general. This limitation can restrict the depth and effectiveness of user engagement. Thus, one option for further work would be to sample more frames for visual augmentation, and include time stamps with them.

To some extent, the wording of the LLM's responses can influence user engagement. In our PoC, it occasionally asks users whether they would like to know more about a particular aspect of the topic under discussion, but this remains rare and somewhat generic. Through careful priming of the LLM (in the prompt we include with every user query), it might be possible to get the LLM to ask more specific questions, either to check that users understood the answer, or to encourage deeper engagement by bringing their attention to other aspects of the dataset. It would also be beneficial for the LLM to ask for clarification when user queries are too ambiguous, to prevent it from responding off topic, or hallucinating.

Currently, we always use the same prompt when sending an image to OpenAI's VISION model, regardless of the dataset. We engineered it to extract visual information effectively, but it does not account for the specificities of each dataset. It might be useful to submit this base prompt to an LLM first, along with the structured JSON description of the dataset, and ask the LLM to adapt the prompt to the dataset. This could make extracting visual information more effective and relevant to each dataset, while keeping the process fully automated.

\subsection{Proof of Concept Limitations}
Our proof of concept also has limitations of its own. Some are linked to the method's limitation. For example, our PoC is unable to generate or modify visualizations dynamically. It relies exclusively on pre-rendered content, which limits the system's interactivity and responsiveness to user inquiries. Unlike a human facilitator who can effortlessly point to or highlight specific visualization parts in real time, our assistant cannot adapt the visual content on the fly to address spontaneous questions or emphasize particular elements. This is related to the LLM's ignorance of the visualization's exact state at any given time.

But other limitations aren't directly linked to our augmentation method or to the LLM itself. For instance, people tend not to spontaneously strike up a conversation with our LLM assistant unless encouraged to do so by a human introducing them to the system. This is unsurprising, given that our PoC does not feature an embodied agent or implement any form of management of social cues~\cite{feine2019taxonomy}. Still, we acknowledge that a fully human-like avatar is not strictly necessary to foster meaningful interactions. Abstract forms of embodiment, such as virtual agents represented through geometric shapes or stylized figures could still provide a sense of presence and facilitate engagement without the complexity of simulating human appearance and behavior.

From an educational perspective, our PoC is limited by its inability to sense user confusion or loss of focus, or to process multi-modal user inputs beyond text and voice. Comparatively, human facilitators excel at interpreting various social cues---gestures, facial expressions, and different voice tones---that enhance interactions and provide valuable context. These capabilities allow them to detect when a user is confused or disengaged and adjust their approach to maintain effective learning experiences. Our assistant, however, relies primarily on textual input and cannot interpret nonverbal signals: we transcribe the user's input voice, potentially leading to less nuanced and effective communication, as even the user's tone is lost. Addressing this limitation would likely improve user engagement and learning outcomes. Efforts to tackle these challenges have been made in prior work~\cite{behera2020associating, tang2015automatic, tang2024facial}.

Future work will focus on augmenting the LLM with time-dependent visual information, and enabling dynamic generation and modification of visualizations for real-time adaptability. It should also incorporate abstract embodiment like virtual agents to enrich engagement. Efforts should be made to improve the prompts sent to the VISION model, and make the LLM's responses more engaging. We also intend to investigate the quality and latency that other LLMs can provide for user queries, particularly very low-cost options like GPT-4o-mini,\footnotemark{} though we expect that always using the best available LLM for all pre-processing steps would be sensible, since this only needs to be performed once per dataset, regardless of the number of users or of the number of queries per user.

\footnotetext{\footnotesize{\url{https://platform.openai.com/docs/models/gpt-4o-mini}}}

\section{Conclusion}

We have developed a novel method for augmenting an LLM with a coherent combination of text and visual data, relevant to a specific topic. This augmentation uses a clear, structured format that is highly compact, improving reliability, and saving on tokens, therefore reducing costs during API calls to the LLM. Our approach does not involve any fine-tuning, so that new data can quickly and easily be used for further augmentation. We have implemented it in a proof of concept for the visualization of geospatial data from the \SoS project. Through an evaluation process comparing it to pure text augmentation and to full text plus vision augmentation (but without the use of a structured format), we show that our approach significantly outperforms both, perhaps surprisingly in the case of text plus vision augmentation, since our approach augments the LLM with a \myemph{lower} amount of data. Thus, we also achieve this with fewer tokens per prompt and, therefore, lower cost per call to the API.

We have shown our proof of concept to be well-liked by our participants, particularly when we consider \myemph{System Usability Scale} items that are most relevant to global \myemph{TellUs} project goals. We have also identified key limitations in our work and proposed several axes for future investigation, laying the groundwork for more comprehensive visual augmentation of LLMs, and for the full realization of the \myemph{TellUs} project.

\section*{Acknowledgments}
The work was supported in part by the Knut and Alice Wallenberg Foundation (grant KAW 2019.0024), the National Science Foundation (US) award 2007436 and by the
King Abdullah University of Science and Technology (BAS/1/1680-01-01).

\bibliographystyle{IEEEtran}
\bibliography{bibliograph}

\begin{thebibliography}{10}
\providecommand{\url}[1]{#1}
\csname url@samestyle\endcsname
\providecommand{\newblock}{\relax}
\providecommand{\bibinfo}[2]{#2}
\providecommand{\BIBentrySTDinterwordspacing}{\spaceskip=0pt\relax}
\providecommand{\BIBentryALTinterwordstretchfactor}{4}
\providecommand{\BIBentryALTinterwordspacing}{\spaceskip=\fontdimen2\font plus
\BIBentryALTinterwordstretchfactor\fontdimen3\font minus \fontdimen4\font\relax}
\providecommand{\BIBforeignlanguage}[2]{{%
\expandafter\ifx\csname l@#1\endcsname\relax
\typeout{** WARNING: IEEEtran.bst: No hyphenation pattern has been}%
\typeout{** loaded for the language `#1'. Using the pattern for}%
\typeout{** the default language instead.}%
\else
\language=\csname l@#1\endcsname
\fi
#2}}
\providecommand{\BIBdecl}{\relax}
\BIBdecl

\bibitem{smith2011aesthetics}
L.~F. Smith, J.~K. Smith, K.~K. Arcand, R.~K. Smith, J.~Bookbinder, and K.~Keach, ``Aesthetics and astronomy: Studying the public’s perception and understanding of imagery from space,'' \emph{Science Communication}, vol.~33, no.~2, pp. 201--238, 2011.

\bibitem{goldman2010science}
K.~H. Goldman, C.~Kessler, and E.~Danter, ``Science on a sphere{\textregistered},'' \emph{Retrieved December}, vol.~31, p. 2018, 2010.

\bibitem{tellus}
L.~Besançon, M.~Brossier, O.~Mena, E.~Sundén, A.~Göransson, A.~Ynnerman, and K.~J. Schönborn, ``Tellus – {L}everaging the power of {LLM}s with visualization to benefit science centers.'' \emph{Nightingale Magazine}, 2024, to appear.

\bibitem{doerner2022virtual}
R.~Doerner, W.~Broll, P.~Grimm, and B.~Jung, \emph{Virtual and augmented reality (VR/AR): Foundations and methods of extended realities (XR)}.\hskip 1em plus 0.5em minus 0.4em\relax Springer Nature, 2022.

\bibitem{peng2023check}
B.~Peng, M.~Galley, P.~He, H.~Cheng, Y.~Xie, Y.~Hu, Q.~Huang, L.~Liden, Z.~Yu, W.~Chen \emph{et~al.}, ``Check your facts and try again: Improving large language models with external knowledge and automated feedback,'' \emph{arXiv preprint arXiv:2302.12813}, 2023.

\bibitem{narechania2020nl4dv}
A.~Narechania, A.~Srinivasan, and J.~Stasko, ``Nl4dv: A toolkit for generating analytic specifications for data visualization from natural language queries,'' \emph{IEEE Transactions on Visualization and Computer Graphics}, vol.~27, no.~2, pp. 369--379, 2020.

\bibitem{maddigan2023chat2vis}
P.~Maddigan and T.~Susnjak, ``Chat2vis: generating data visualizations via natural language using chatgpt, codex and gpt-3 large language models,'' \emph{IEEE Access}, vol.~11, pp. 45\,181--45\,193, 2023.

\bibitem{liu2021advisor}
C.~Liu, Y.~Han, R.~Jiang, and X.~Yuan, ``Advisor: Automatic visualization answer for natural-language question on tabular data,'' in \emph{2021 IEEE 14th Pacific Visualization Symposium (PacificVis)}.\hskip 1em plus 0.5em minus 0.4em\relax IEEE, 2021, pp. 11--20.

\bibitem{devlin2018bert}
J.~Devlin, ``Bert: Pre-training of deep bidirectional transformers for language understanding,'' \emph{arXiv preprint arXiv:1810.04805}, 2018.

\bibitem{chen2024viseval}
N.~Chen, Y.~Zhang, J.~Xu, K.~Ren, and Y.~Yang, ``Viseval: A benchmark for data visualization in the era of large language models,'' \emph{IEEE Transactions on Visualization and Computer Graphics}, 2024.

\bibitem{mitra2022facilitating}
R.~Mitra, A.~Narechania, A.~Endert, and J.~Stasko, ``Facilitating conversational interaction in natural language interfaces for visualization,'' in \emph{2022 IEEE Visualization and Visual Analytics (VIS)}.\hskip 1em plus 0.5em minus 0.4em\relax IEEE, 2022, pp. 6--10.

\bibitem{satyanarayan2016vega}
A.~Satyanarayan, D.~Moritz, K.~Wongsuphasawat, and J.~Heer, ``Vega-lite: A grammar of interactive graphics,'' \emph{IEEE transactions on visualization and computer graphics}, vol.~23, no.~1, pp. 341--350, 2016.

\bibitem{voigt2023vist5}
H.~Voigt, N.~Carvalhais, M.~Meuschke, M.~Reichstein, S.~Zarrie, and K.~Lawonn, ``Vist5: An adaptive, retrieval-augmented language model for visualization-oriented dialog,'' in \emph{The 2023 Conference on Empirical Methods in Natural Language Processing}.\hskip 1em plus 0.5em minus 0.4em\relax Association for Computational Linguistics, 2023, pp. 70--81.

\bibitem{hong2023conversational}
M.-H. Hong and A.~Crisan, ``Conversational ai threads for visualizing multidimensional datasets,'' \emph{arXiv preprint arXiv:2311.05590}, 2023.

\bibitem{jia2023voice}
D.~Jia, A.~Irger, L.~Besancon, O.~Strnad, D.~Luo, J.~Bjorklund, A.~Ynnerman, and I.~Viola, ``Voice: Visual oracle for interaction, conversation, and explanation,'' \emph{arXiv preprint arXiv:2304.04083}, 2023.

\bibitem{kafle2018dvqa}
K.~Kafle, B.~Price, S.~Cohen, and C.~Kanan, ``Dvqa: Understanding data visualizations via question answering,'' in \emph{Proceedings of the IEEE conference on computer vision and pattern recognition}, 2018, pp. 5648--5656.

\bibitem{masry2022chartqa}
A.~Masry, D.~X. Long, J.~Q. Tan, S.~Joty, and E.~Hoque, ``Chart{QA}: A benchmark for question answering about charts with visual and logical reasoning,'' \emph{arXiv preprint arXiv:2203.10244}, 2022.

\bibitem{vaswani2017attention}
A.~Vaswani, ``Attention is all you need,'' \emph{Advances in Neural Information Processing Systems}, 2017.

\bibitem{dosovitskiy2020image}
A.~Dosovitskiy, ``An image is worth 16x16 words: Transformers for image recognition at scale,'' \emph{arXiv preprint arXiv:2010.11929}, 2020.

\bibitem{li2021align}
J.~Li, R.~Selvaraju, A.~Gotmare, S.~Joty, C.~Xiong, and S.~C.~H. Hoi, ``Align before fuse: Vision and language representation learning with momentum distillation,'' \emph{Advances in neural information processing systems}, vol.~34, pp. 9694--9705, 2021.

\bibitem{bendeck2024empirical}
A.~Bendeck and J.~Stasko, ``An empirical evaluation of the gpt-4 multimodal language model on visualization literacy tasks,'' \emph{IEEE Transactions on Visualization and Computer Graphics}, 2024.

\bibitem{feng2023geoqamap}
Y.~Feng, L.~Ding, and G.~Xiao, ``Geoqamap-geographic question answering with maps leveraging llm and open knowledge base (short paper),'' in \emph{12th International Conference on Geographic Information Science (GIScience 2023)}.\hskip 1em plus 0.5em minus 0.4em\relax Schloss Dagstuhl-Leibniz-Zentrum f{\"u}r Informatik, 2023.

\bibitem{harris2013sparql}
S.~Harris, ``Sparql 1. 1 query language,'' \emph{W3C Recommendation}, vol.~21, 2013.

\bibitem{chang2022mapqa}
S.~Chang, D.~Palzer, J.~Li, E.~Fosler-Lussier, and N.~Xiao, ``Mapqa: A dataset for question answering on choropleth maps,'' \emph{arXiv preprint arXiv:2211.08545}, 2022.

\bibitem{chartsAsText}
V.~S. Bursztyn, J.~Hoffswell, E.~Koh, and S.~Guo, ``Representing charts as text for language models: An in-depth study of question answering for bar charts,'' \emph{IEEE Transactions on Visualization and Computer Graphics}, 2024.

\bibitem{young2013pomdp}
S.~Young, M.~Ga{\v{s}}i{\'c}, B.~Thomson, and J.~D. Williams, ``Pomdp-based statistical spoken dialog systems: A review,'' \emph{Proceedings of the IEEE}, vol. 101, no.~5, pp. 1160--1179, 2013.

\bibitem{patel2021projected}
S.~H. Patel, M.~V. Winton, J.~M. Hatch, H.~L. Haas, V.~S. Saba, G.~Fay, and R.~J. Smolowitz, ``Projected shifts in loggerhead sea turtle thermal habitat in the northwest atlantic ocean due to climate change,'' \emph{Scientific Reports}, vol.~11, no.~1, p. 8850, 2021.

\bibitem{brown2020language}
T.~B. Brown, ``Language models are few-shot learners,'' \emph{arXiv preprint arXiv:2005.14165}, 2020.

\bibitem{shiftsTurtles}
S.~Patel, M.~Winton, J.~Hatch, H.~Haas, V.~Saba, G.~Fay, and R.~Smolowitz, ``Projected shifts in loggerhead sea turtle habitat in the northwest atlantic ocean due to climate change,'' 12 2020.

\bibitem{bangor2008empirical}
A.~Bangor, P.~T. Kortum, and J.~T. Miller, ``An empirical evaluation of the system usability scale,'' \emph{Intl. Journal of Human--Computer Interaction}, vol.~24, no.~6, pp. 574--594, 2008.

\bibitem{brooke1996sus}
J.~Brooke, ``Sus: A quick and dirty usability scale,'' \emph{Usability Evaluation in Industry}, 1996.

\bibitem{shapiro1965analysis}
S.~S. Shapiro and M.~B. Wilk, ``An analysis of variance test for normality (complete samples),'' \emph{Biometrika}, vol.~52, no. 3-4, pp. 591--611, 1965.

\bibitem{hollander2013nonparametric}
M.~Hollander, D.~A. Wolfe, and E.~Chicken, \emph{Nonparametric statistical methods}.\hskip 1em plus 0.5em minus 0.4em\relax John Wiley \& Sons, 2013.

\bibitem{rupert2012simultaneous}
G.~Rupert~Jr \emph{et~al.}, ``Simultaneous statistical inference,'' 2012.

\bibitem{feine2019taxonomy}
J.~Feine, U.~Gnewuch, S.~Morana, and A.~Maedche, ``A taxonomy of social cues for conversational agents,'' \emph{International Journal of human-computer studies}, vol. 132, pp. 138--161, 2019.

\bibitem{behera2020associating}
A.~Behera, P.~Matthew, A.~Keidel, P.~Vangorp, H.~Fang, and S.~Canning, ``Associating facial expressions and upper-body gestures with learning tasks for enhancing intelligent tutoring systems,'' \emph{International Journal of Artificial Intelligence in Education}, vol.~30, pp. 236--270, 2020.

\bibitem{tang2015automatic}
C.~Tang, P.~Xu, Z.~Luo, G.~Zhao, and T.~Zou, ``Automatic facial expression analysis of students in teaching environments,'' in \emph{Biometric Recognition: 10th Chinese Conference, CCBR 2015, Tianjin, China, November 13-15, 2015, Proceedings 10}.\hskip 1em plus 0.5em minus 0.4em\relax Springer, 2015, pp. 439--447.

\bibitem{tang2024facial}
X.~Tang, Y.~Gong, Y.~Xiao, J.~Xiong, and L.~Bao, ``Facial expression recognition for probing students’ emotional engagement in science learning,'' \emph{Journal of Science Education and Technology}, pp. 1--18, 2024.

\end{thebibliography}

\begin{IEEEbiography}
[{\includegraphics[width=1in,height=1.25in,clip,keepaspectratio]{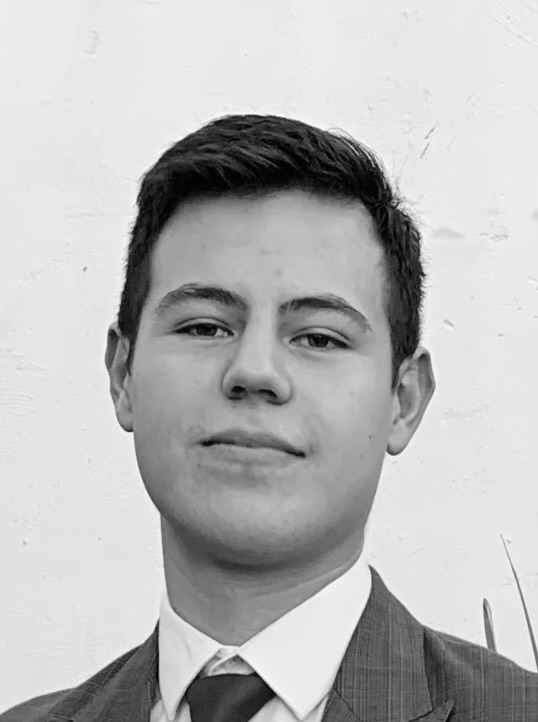}}]{Omar Mena}
is currently a MSc student in Computer Science at King Abdullah University of Science and Technology (KAUST), Saudi Arabia. He obtained his BSc in Artificial Intelligence at Universidad Panamericana, Mexico. His research interests are XR computer-human interactions and computer graphics.
\end{IEEEbiography}

\begin{IEEEbiography}[{\includegraphics[width=1in,height=1.25in,clip,keepaspectratio]{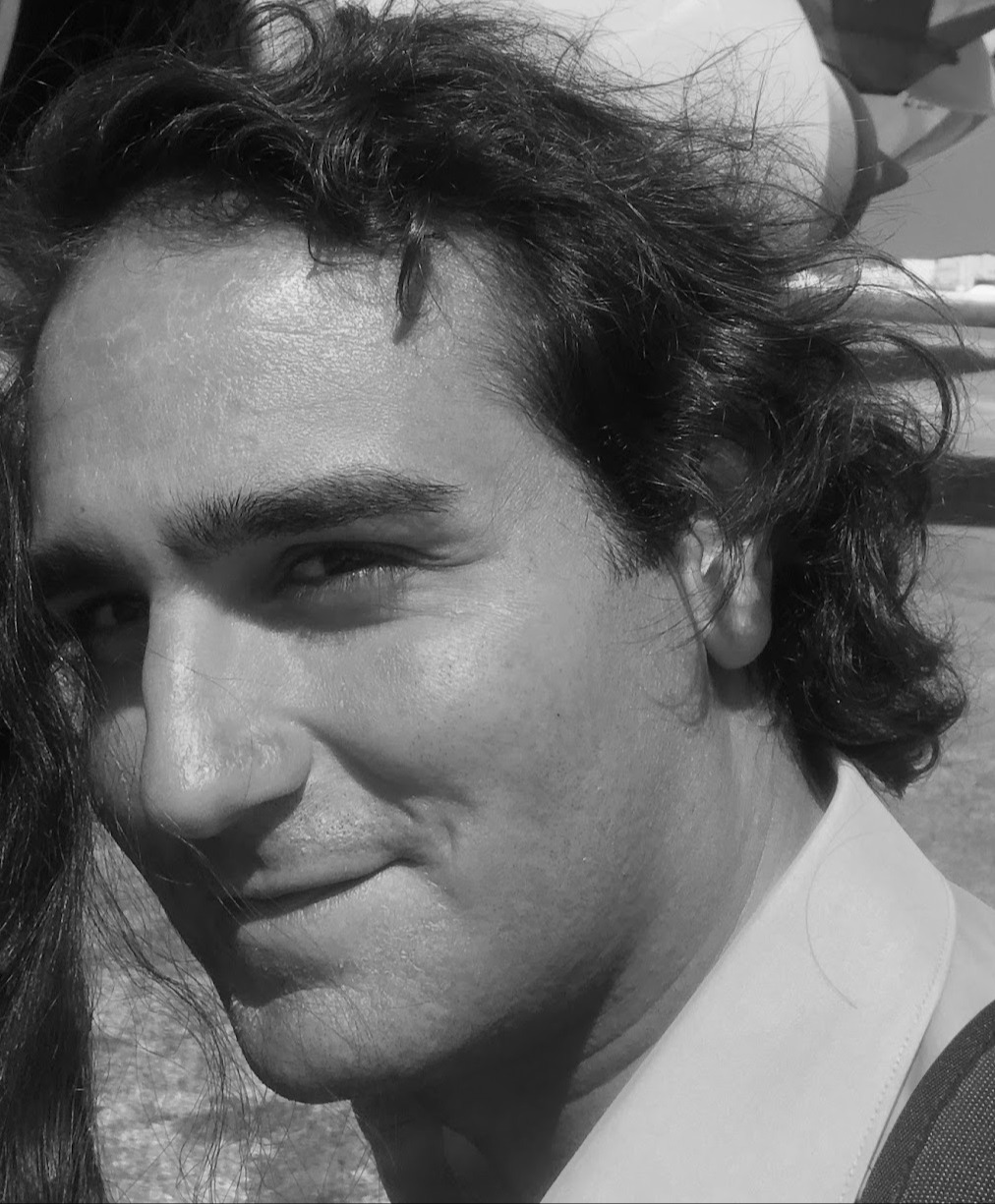}}]{Alexandre Kouyoumdjian}
is a research scientist at King Abdullah University of Science and Technology (KAUST). He holds a PhD in computer science from University Paris-Saclay. He conducts research on multiscale visualization, interaction, and modeling for biology, with an additional focus on virtual and augmented reality.
\end{IEEEbiography}

\begin{IEEEbiography}[{\includegraphics[width=1in,height=1.25in,clip,keepaspectratio]{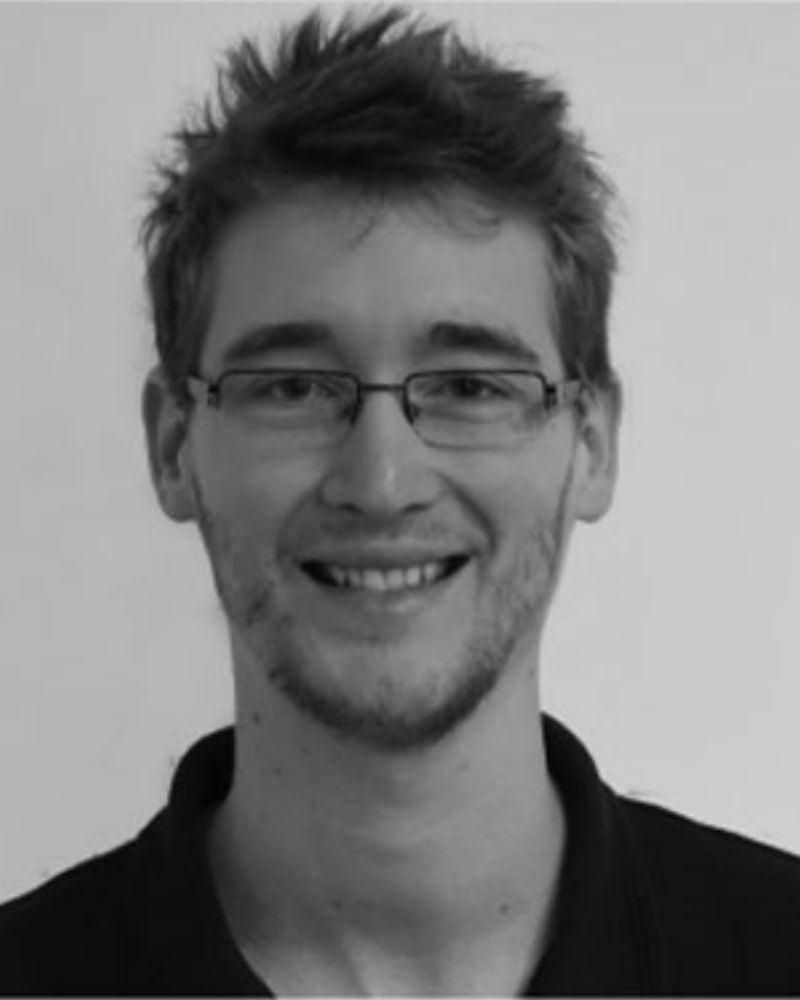}}]{Lonni Besançon}
received a PhD degree in computer science from the University Paris-Saclay, France. He is currently an assistant professor with Linköping University, Sweden. His research focuses on interactive visualization techniques for 3D spatial data relying on new input paradigms, which include the visualization and understanding of uncertainty in empirical results in computer science.
\end{IEEEbiography}

\begin{IEEEbiography}[{\includegraphics[width=1in,height=1.25in,clip,keepaspectratio]{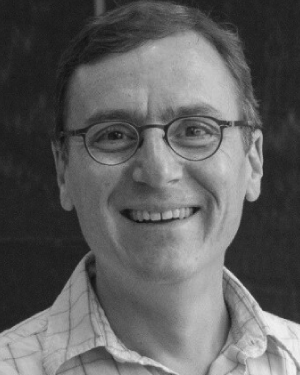}}]{Michael Gleicher}
is a Professor in the Department of Computer Sciences at the University of Wisconsin, Madison. Prof. Gleicher is founder of the Department’s Visual Computing Group and co-directs both the Visual Computing Laboratory and the Collaborative Robotics Laboratory at UW-Madison. His research interests span the range of visual computing, including data visualization, robotics, and virtual/extended reality. His recent work includes exploring perceptual issues in visualization, the use of visual simulation for robotics, and geometric approaches to enhance robot perception and interaction. He earned his Ph. D. in Computer Science (1994) from Carnegie Mellon University, and earned a B.S.E. in Electrical Engineering from Duke University (1988). In 2023-2024, Prof. Gleicher holds a concurrent appointment as a Design Scholar at Amazon Robotics. This work is not associated with Amazon.
\end{IEEEbiography}

\begin{IEEEbiography}[{\includegraphics[width=1in,height=1.25in,clip,keepaspectratio]{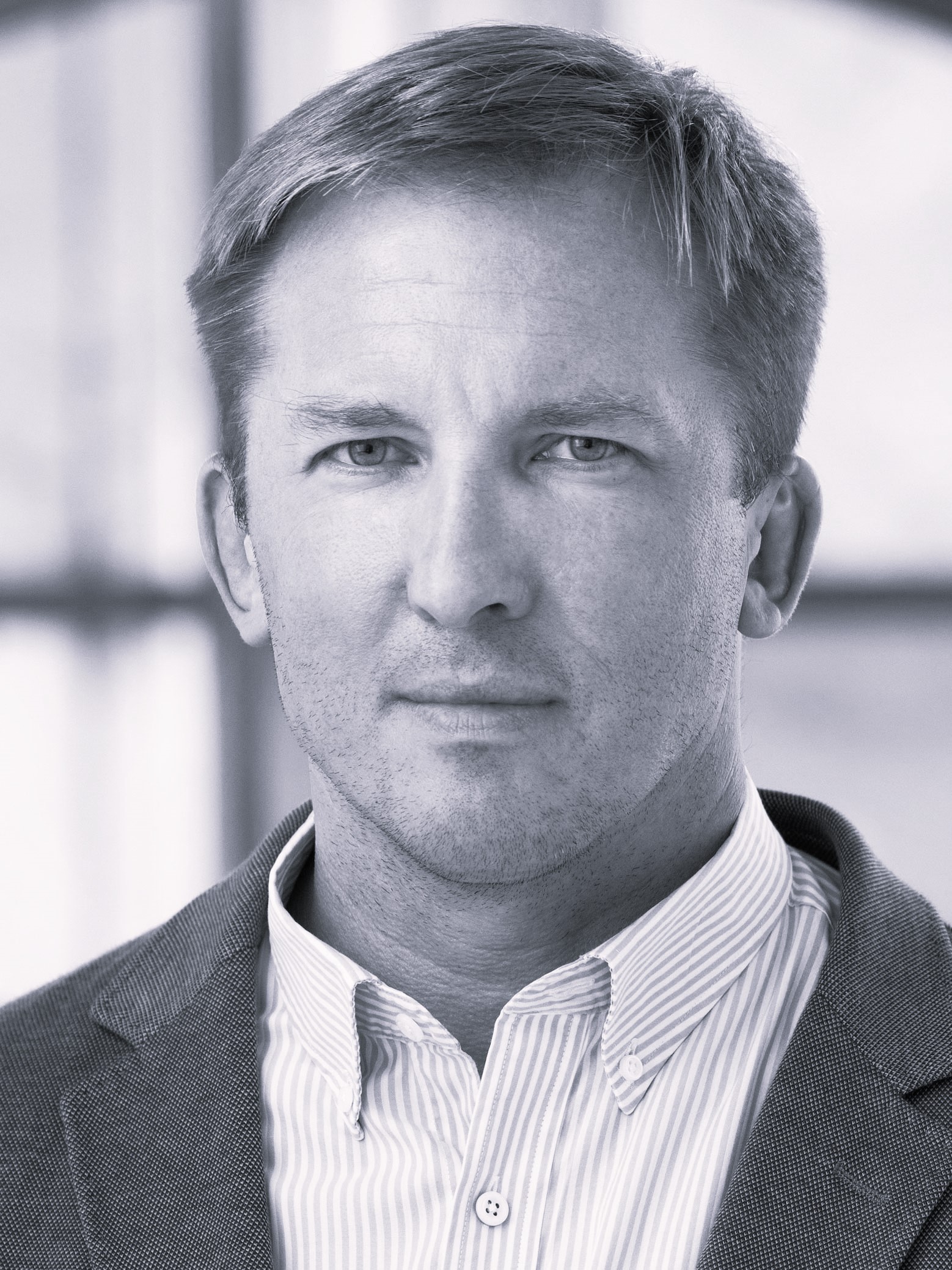}}]{Ivan Viola}
is a professor at King Abdullah University of Science and Technology (KAUST), Saudi Arabia. He graduated from TU Wien, Austria. In 2005 he took a postdoc position at the University of Bergen, Norway, where he was gradually promoted to the professor rank. In 2013 he received a WWTF grant to establish a research group at TU Wien. At KAUST, he continues developing new visualization techniques, primarily oriented on data reconstruction, interpretation, representation, modeling, and rendering.
\end{IEEEbiography}

\begin{IEEEbiography}[{\includegraphics[width=1in,height=1.25in,clip,keepaspectratio]{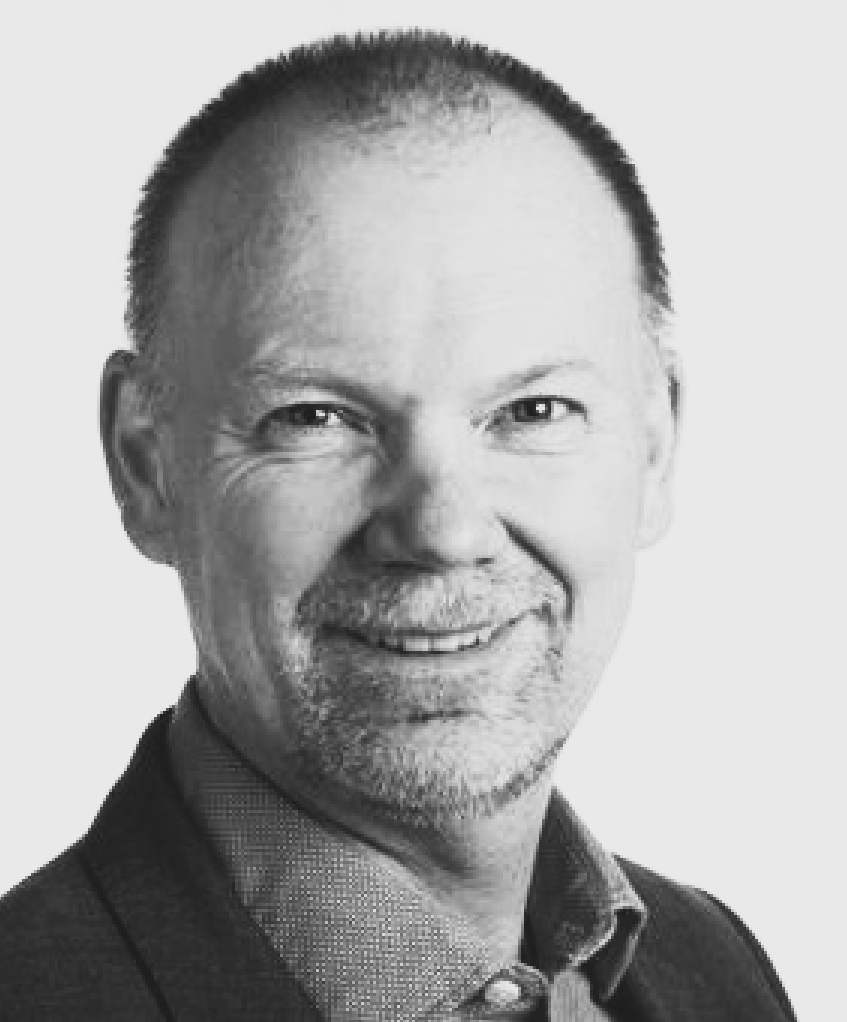}}]{Anders Ynnerman}
(Member, IEEE) is currently a
professor in scientific visualization with Linköping University and the director of Norrköping Visualization Center C. His research focuses on interactive
techniques for large scientific data in a range of application areas. In 2018, he was the recipient of the IEEE VGTC Technical Achievement Award. He is currently a member of the Swedish Royal Academy of Engineering Sciences and the Royal Swedish Academy of Sciences.
\end{IEEEbiography}

\vfill

\end{document}